\documentclass[aps,prb,twocolumn,amsmath,superscriptaddress]{revtex4-2}

\usepackage{graphicx}
\usepackage{color}
\usepackage{epstopdf}
\usepackage{xcolor}

\begin{document}

\title{Spin dynamics of positively charged excitons in Cr$^+$-doped quantum dots probed by resonant photoluminescence.}

\author{V.~Tiwari}
\affiliation{Institut N\'{e}el, CNRS, Univ. Grenoble Alpes and Grenoble INP, 38000 Grenoble, France}

\author{M.~Morita}
\affiliation{Institute of Materials Science, University of Tsukuba, 1-1-1 Tennoudai, Tsukuba, 305-8573, Japan}

\author{T.~Inoue}
\affiliation{Institute of Materials Science, University of Tsukuba, 1-1-1 Tennoudai, Tsukuba, 305-8573, Japan}

\author{S.~Ando}
\affiliation{Institute of Materials Science, University of Tsukuba, 1-1-1 Tennoudai, Tsukuba, 305-8573, Japan}

\author{S.~Kuroda}
\affiliation{Institute of Materials Science, University of Tsukuba, 1-1-1 Tennoudai, Tsukuba, 305-8573, Japan}

\author{H.~Boukari}
\affiliation{Institut N\'{e}el, CNRS, Univ. Grenoble Alpes and Grenoble INP, 38000 Grenoble, France}

\author{L.~Besombes}\email{lucien.besombes@neel.cnrs.fr}
\affiliation{Institut N\'{e}el, CNRS, Univ. Grenoble Alpes and Grenoble INP, 38000 Grenoble, France}

\date{\today}

\begin{abstract}

We study the dynamics of the spin system that consist of a positively charged II-VI semiconductor quantum dot doped with a single Cr$^+$ ion. The resonant photoluminescence (PL) of the positively charged exciton coupled with the Cr$^+$ spin is used to analyze the main spin relaxation channels. The intensity of the resonant PL is reduced by an optical pumping of the spin of the resident hole-Cr$^+$ complex that can be seen as a nano-magnet. The spin memory can be partially erased by a non-resonant optical excitation. This leads to an increase of the resonant PL signal. The resonant PL is co-circularly polarized and corresponds to relaxation channels that conserve the Cr$^+$ spin $\vert S_z \vert$. The observation in the resonant-PL excitation spectra of optical transitions with a change of the Cr$^+$ spin permits to determine the magnetic anisotropy of the magnetic atom. Optical pumping, auto-correlation measurements and the power dependence of the PL intensity distribution show that the effective temperature of the hole-Cr$^+$ spin system is affected by the optical excitation through the local generation of phonons. 

\end{abstract}

\maketitle

\section{Introduction}

Point defects in solid-sate materials have emerged as an important platform for quantum technology applications. Individual localized spins in semiconductors have already found some applications in defect based quantum systems \cite{Wolfowicz2021}. Many point defects exist in semiconductors and a detail understanding of their properties (charge, spin, optics, mechanics, …) and their interaction with the host material will permit discovering new relevant systems for specific quantum applications. Among these defects, magnetic transition metal atoms incorporated in conventional semiconductors are of particular interest as they offer a large choice of localized electronic spins, nuclear spins as well as orbital momentum \cite{Besombes2004,Kudelski2007,Kobak2014}. The exchange interaction of the spin localized on the atom with the spin of the free carriers of the semiconductor host also permit to use the optical or electrical properties of semiconductors to interact with the magnetic atoms. In particular, the spin of individual magnetic atoms can be probed and controlled through their interaction with confined carriers in a quantum dot (QD) \cite{Besombes2012,Krebs2013,Varghese2014,Bacher2016,Bacher2020,Bayer2020}. Inserting magnetic atoms in QDs allows to engineer the spin properties of the atom through a control of the electric charge or the strain \cite{Besombes2014, Lafuente2017}. In addition to carrying localized spins, transition metal elements can also be electrical dopants offering an additional possibility of electrical control. This is in particular the case of chromium (Cr) in II-VI semiconductor compounds \cite{Ciepla1975,Godlewski1980}. 

Cr is naturally incorporated in II-VI semiconductor compounds as a Cr$^{2+}$ ion. Cr$^{2+}$ is the isoelectronic configuration of Cr in the II-VI semiconductor lattice. It has been demonstrated recently that the Cr in its Cr$^+$ oxidation state can also be stable and detected optically when inserted in a CdTe QD \cite{Tiwari2021}. The Cr$^+$ has an electronic configuration 3$d^5$ with a spin S=5/2 and no orbital momentum (L=0). In this case, the Cr atom is a negatively charged defect in the semiconductor lattice. It localizes a heavy-hole (angular momentum J$_z=$+3/2 ($\Uparrow_h$) or J$_z=$-3/2 ($\Downarrow_h$)) in the QD and a ferromagnetic hole-Cr$^+$ complex is formed with a quantification axis along the QD growth axis. This complex can be seen as a nano-magnet with two ground states having a total angular momentum $M_z=\pm 4$ with parallel hole and Cr$^+$ spins.

In optically active QDs, resonant photoluminescence (PL) and optical spin pumping are the most efficient tools to initialize and probe the dynamics of individual spins. It has been demonstrated that resonant optical pumping can be used to empty the low energy states of the hole-Cr$^+$ complex with a spin memory in the 20 $\mu s$ range at low temperature \cite{Tiwari2021}. The spin dynamics of hole-Cr$^+$ was found to depend on the conditions of optical excitation. The spin memory of the hole-Cr$^+$ complex also limits the PL that can be obtained under resonant excitation of the positively charged exciton (X$^+$-Cr$^+$). To reveal the main spin-flips channels among X$^+$-Cr$^+$ that can be involved in the optical pumping of a hole-Cr$^+$ complex we analyze here the structure of the PL of X$^+$-Cr$^+$ and its dynamics under a tunable resonant excitation.

The article is organized as follows: After a short presentation of the samples and the experiments (section II) we recall in section III the energy level structure in positively charged Cr$^+$-doped QDs and its dependence in low magnetic field. We show in particular that optical transitions with a change of the Cr$^+$ spin are possible. In section IV we present the structure of the resonant PL of X$^+$-Cr$^+$. We show first that its intensity is influenced by an additional non-resonant excitation. We demonstrate that the main channels of spin transfer within X$^+$-Cr$^+$ under resonant excitation correspond to a spin tunneling. Optical transitions that do not conserve the Cr$^+$ spin are also observed in the resonant PL excitation spectra. Their energy positions are controlled by the magnetic anisotropy of the Cr$^+$. In section V we demonstrate that the resonant PL is reduced by the optical pumping of the hole-Cr$^+$ spin and show that an optical pumping of the hole-Cr$^+$ complex can occur for an excitation on the acoustic phonon side bands of the X$^+$-Cr$^+$ lines. At high resonant excitation power, the resonant pumping is perturbed by the generation of acoustic phonons (section VI). A summary and discussions are presented in section VII.

\section{Samples and experiments}

In this study we use self-assembled CdTe/ZnTe QDs grown by molecular beam epitaxy and doped with Cr atoms. The concentration of Cr was adjusted to obtain QDs containing 0, 1 or a few magnetic atoms. When doped with a single atom these dots permit to optically access the spin of a Cr$^{2+}$ ion \cite{Lafuente2016} or, in the presence of a local background doping, to the spin of a hole-Cr$^+$ complex \cite{Tiwari2021}. 

Individual QDs were studied by optical micro-spectroscopy at liquid helium temperature (T=4.2 K). The PL was excited with a tunable dye laser, collected by a high numerical aperture microscope objective (NA=0.85), dispersed by a two meters double monochromator and detected by a cooled Si CCD. For PL measurements, the sample was inserted in the bore of a superconducting coil and a magnetic field could be applied along the QD growth axis. For time resolved optical pumping experiments a dye laser was tuned on resonance with one of the optical transition of X$^+$-Cr$^+$. The laser power was stabilized by an electro-optic variable attenuator. A second non-resonant laser at a wavelength of 568 nm, far below the absorption edge of the ZnTe barriers, was used for a high energy  non spin selective excitation. Both lasers could be modulated by acoustic-optic modulators with a rise and fall time of about 10 ns. A Si avalanche photodiode (APD) with a time resolution of about 350 ps combined with a time-resolved photon counting system was used for the detection.

\section{Energy levels in positively charged Cr$^+$-doped QDs.}

\begin{figure}[hbt]
\centering
\includegraphics[width=1.0\linewidth]{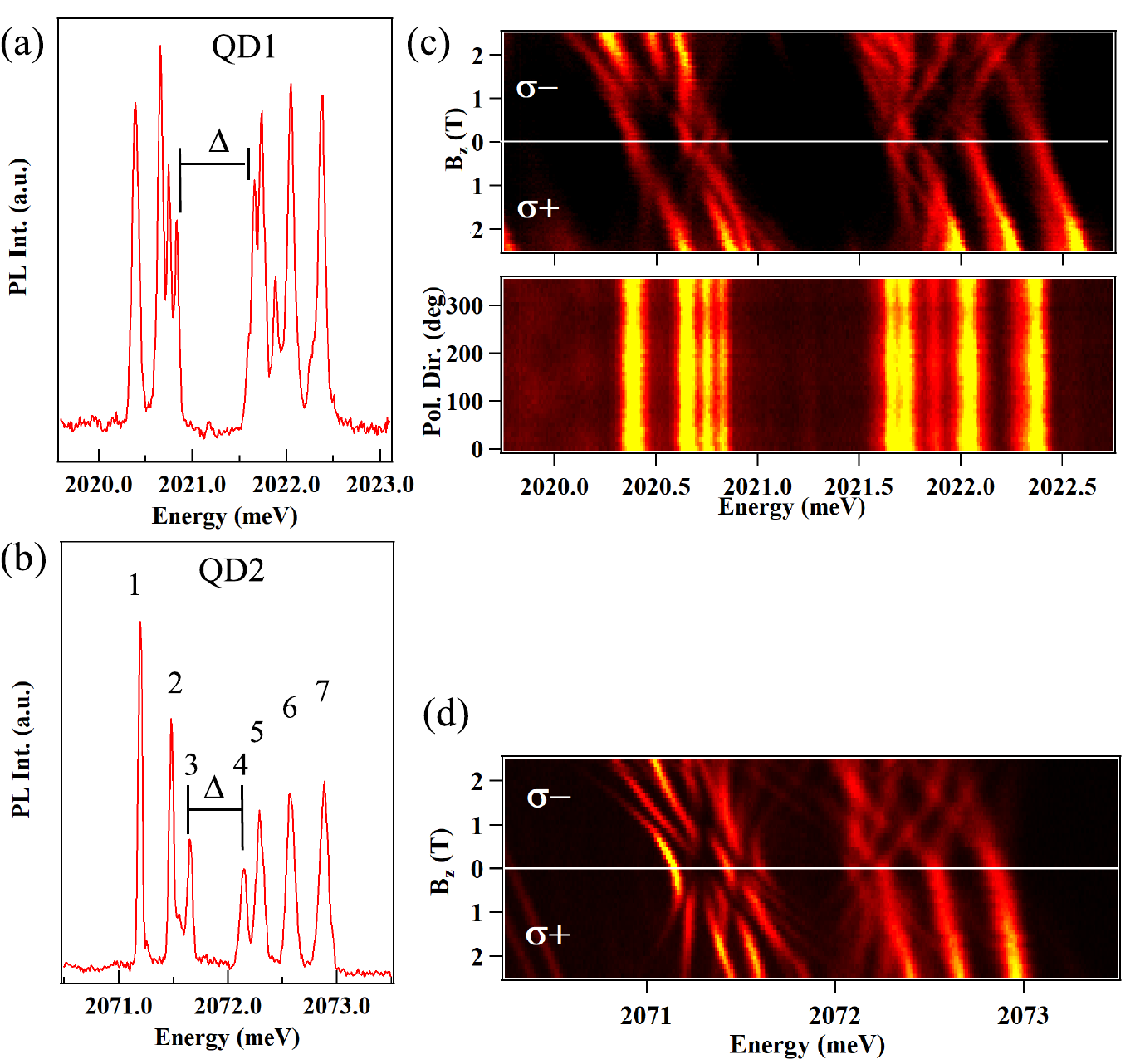}
\caption{Low temperature (T=5K) PL of two Cr$^+$-doped QDs, QD1 (a) and QD2 (b). (c) Intensity map of the longitudinal magnetic field dependence (top panel) and of the linear polarization dependence (bottom panel) of the PL of QD1. (d) Intensity map of the magnetic field dependence of the PL of QD2.}
\label{FigLevels}
\end{figure}

To understand the resonant PL of a positively charged Cr$^+$-doped QD and the spin dynamics of X$^+$-Cr$^+$, we first analyze the energy level structure in these magnetic QDs. The low temperature PL of two positively charged Cr$^+$-doped QDs are presented in Fig.\ref{FigLevels} together with their low magnetic field dependence in a Faraday configuration. The emission of such dots at zero magnetic field consists in a minimum of 7 lines separated by a central gap $\Delta$ \cite{Tiwari2021}. Additional lower intensity lines can be observed in some of the dots (see QD1). Even in the presence of these additional lines, no linearly polarized split lines are observed at zero magnetic field (Fig.\ref{FigLevels}(c)). This is in striking difference with positively charged QDs doped with Mn$^{2+}$, the most studied 3$d^5$ (S=5/2) diluted magnetic semiconductor system. In the latter case, linearly polarized lines arising from the recombination of the coupled X$^+$-Mn$^{2+}$ states towards hole-Mn$^{2+}$ levels affected by a valence band mixing are characteristic of the X$^+$-Mn$^{2+}$ PL \cite{Lafuente2015}.

The energy levels of the excited state in a Cr$^+$-doped QD, X$^+$-Cr$^+$, are described by the Hamiltonian
\begin{eqnarray}
{\cal H}_{X^+-Cr^+}=I_{eCr^+}\vec{S}\cdot\vec{\sigma}+g_{e}\mu_B\vec{\sigma}\cdot\vec{B}+{\cal H}_{Cr^+}
\end{eqnarray}
\noindent which contains a possible exchange interaction between the Cr$^+$ spin ($\vec{S}$) and the electron spin ($\vec{\sigma}$), the Zeeman energy of the electron spin and the Cr$^+$ Hamiltonian 
\begin{eqnarray}
{\cal H}_{Cr^+}&=&D_0S^2_z+E(S_x^2-S_y^2)+\\ \nonumber
               & &g_{Cr^+}\mu_B\vec{S}\cdot\vec{B}
\end{eqnarray}
\noindent describing the fine structure of the Cr$^+$ spin and its Zeeman energy ($g_{Cr^+}\approx2$) \cite{Godlewski1980,Ludwig1963}. The fine structure arises from the crystal field and the spin-orbit coupling \cite{Vallin1974}. The term $D_0$ commonly called "magnetic anisotropy" is induced by biaxial strain in the plane of the QD. An anisotropy of the strain in the QD plane induces a mixing of the Cr$^+$ spin states described by the $E$ term. Further reduction of the symmetry by local strain can induce additional terms \cite{Vallin1974}. We also neglect here possible terms arising from the tetragonal symmetry of the crystal field in the zinc-blend lattice, assumed to be weak in self-assembled QDs compared with the strain induced terms. 

\begin{figure}[hbt]
\centering
\includegraphics[width=1.0\linewidth]{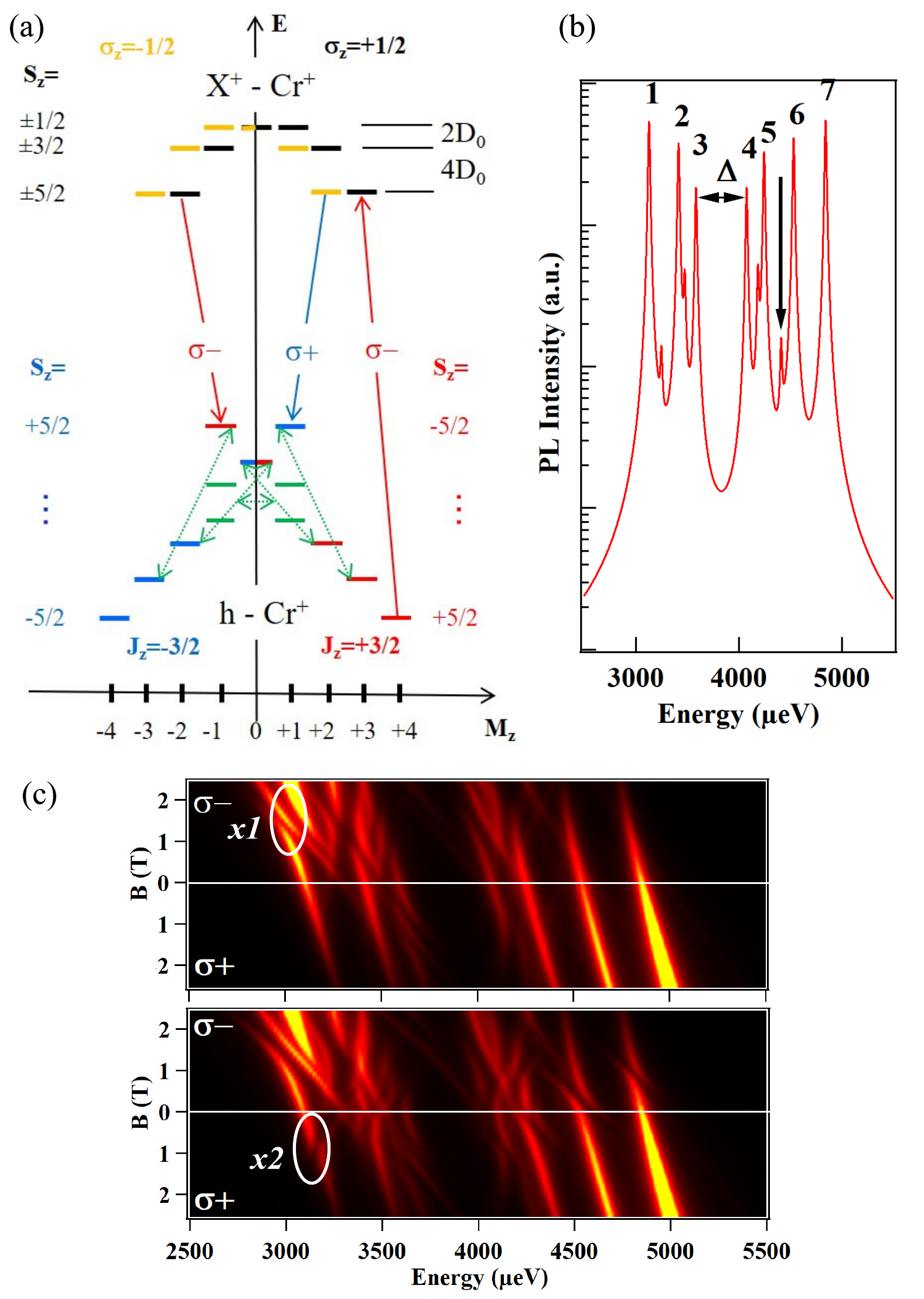}
\caption{(a) Energy levels scheme of the h-Cr$^+$ and the X$^+$-Cr$^+$ states in a positively charged QD containing a Cr$^+$ ion. Within X$^+$-Cr$^+$, the electron-Cr$^+$ exchange interaction and the fine structure term $E$ are neglected. Green arrows correspond to the VBM induced by in-plane strain anisotropy which couples two by two hole-Cr$^+$ levels. (b) Calculated PL spectra (log scale) of X$^+$-Cr$^+$ in the presence of VBM induced by in-plane strain anisotropy $\rho_s/\Delta_{lh}$= 0.21 ($\theta_s$=0) and with the parameters I$_{hCr^+}$= -225 $\mu eV$, I$_{eCr^+}$= 0$\mu eV$, $\xi$=0, $g_{e}$=-0.4, $g_{h}$=0.5, D$_0$=-30 $\mu eV$, E=0 $\mu eV$. (c) Magnetic field dependence of the PL of a Cr$^+$-doped QD calculated with the previous parameters and shear strain $\xi$=0.17 (top panel) and with $\xi$=0.17 and E=10 $\mu eV$ (bottom panel).}
\label{FigMap}
\end{figure}

In Cr$^+$-doped QDs, $I_{eCr^+}\ll D_0$ and the energy levels of X$^+$-Cr$^+$ are dominated in the absence of magnetic field by ${\cal H}_{Cr^+}$. The latter leads in a first approximation to a parabolic splitting of the Cr$^+$ spin states. This structure at zero field is presented in the simplified energy levels scheme of Fig.~\ref{FigMap}(a). In this model, we consider that the exchange interaction of the two spin paired holes with the Cr$^+$ spin is negligible \cite{Lafuente2015, Hawrylak2013} and we also neglect any possible diamagnetic shift of X$^+$ not relevant in the low magnetic field regime.

The hole-Cr$^+$ complex in the ground state is dominated by a ferromagnetic exchange interaction between the hole and the Cr$^+$ spins \cite{Tiwari2021} and the energy levels are described by 
\begin{eqnarray}
{\cal H}_{h-Cr^+}=I_{hCr^+}\vec{S}\cdot\vec{J}+g_{h}\mu_B\vec{J}\cdot\vec{B}+{\cal H}_{Cr^+}
\end{eqnarray}
\noindent where $\vec{J}$ is the hole spin operator and g$_h$ its Landé factor. In the subspace of the two low-energy heavy-hole states, a pseudo-spin operator $\tilde{j}$ can be used to take into account the influence of a possible valence band mixing (VBM). In self-assembled QDs the VBM mainly arises from a strain anisotropy described by the Bir-Pikus Hamiltonian \cite{Leger2007}.

In the presence of anisotropy in the QD plane, the two hole ground states become $\vert\Phi^+_h\rangle=\vert+3/2\rangle-\varphi\vert-1/2\rangle$ and $\vert\Phi^-_h\rangle=\vert-3/2\rangle-\varphi^*\vert+1/2\rangle$ with $\varphi=\rho_s/\Delta_{lh}e^{2i\theta_s}$. $\rho_s$ describes the amplitude of the mixing and $\Delta_{lh}$ the energy splitting of light-holes and heavy-holes induced both by strain and confinement. $\theta_s$ is an angle relative to the [100] axis describing the direction of the in-plane strain anisotropy. A first order development of the angular momentum operator $J$ on the subspace of the perturbed holes $\vert\Phi^{\pm}_h\rangle$ leads to the operators acting on $\left\lbrace +3/2;-3/2\right\rbrace$:
\begin{eqnarray}
\tilde{j}_+= 
\frac{\rho_s}{\Delta_{lh}}\begin{pmatrix}
0 & -2\sqrt{3}e^{-2i\theta_s} \\ 
0 & 0 
\end{pmatrix};\\ \nonumber 
\tilde{j}_-=
\frac{\rho_s}{\Delta_{lh}}\begin{pmatrix}
0 & 0 \\ 
-2\sqrt{3}e^{2i\theta_s} & 0 
\end{pmatrix};
\tilde{j}_z=
\begin{pmatrix}
3/2 & 0 \\ 
0 & -3/2 
\end{pmatrix}
\end{eqnarray}
\noindent $\tilde{j}_+$ and $\tilde{j}_-$ flip the hole spin whereas a measurement of the spin projection along $z$ confirms that they are mainly heavy-holes. This type of VBM unlocks the spin-flips between the hole and its surrounding medium. It allows hole-Cr$^+$ flip-flops and couples the different hole-Cr$^+$ spin states. This is in particular the case for the states $\vert S_z=-1/2,\Uparrow_h\rangle$ and $\vert S_z=+1/2,\Downarrow_h\rangle$ which are initially degenerated and split by the VBM induced mixing. This mixing term is responsible for the opening of the gap $\Delta$ in the center of the X$^+$-Cr$^+$ spectra.

In the presence of shear strain, the two hole ground states become $\vert\Xi^+_h\rangle=\vert+3/2\rangle+\xi\vert+1/2\rangle$ and $\vert\Xi^-_h\rangle=\vert-3/2\rangle-\xi^*\vert-1/2\rangle$. A first order development of the angular momentum operator $J$ on the subspace of the perturbed holes $\vert\Xi^{\pm}_h\rangle$ leads in this case to 
\begin{eqnarray}
\tilde{j}_+= 
\xi\begin{pmatrix}
\sqrt{3} & 0 \\ 
0 & -\sqrt{3} 
\end{pmatrix}; \\ \nonumber
\tilde{j}_-=
\xi^*\begin{pmatrix}
\sqrt{3} & 0 \\ 
0 & -\sqrt{3} 
\end{pmatrix};
\tilde{j}_z=
\begin{pmatrix}
3/2 & 0 \\ 
0 & -3/2 
\end{pmatrix}
\end{eqnarray}
Because of this VBM the hole-Cr$^+$ exchange interaction couples the states $\vert\Xi^{\pm}_h,S_z\rangle$ with the states $\vert\Xi^{\pm}_h,S_z+1\rangle$ and $\vert\Xi^{\pm}_h,S_z-1\rangle$. 

A modeling of the PL of X$^+$-Cr$^+$ and its low longitudinal magnetic field dependence is presented in Fig.~\ref{FigMap}. The overall width of the spectra is controlled by the hole-Cr$^+$ exchange interaction and is given by $3/2 \times 5 \times I_{eCr^+}$. The central gap $\Delta$ arises from the VBM induced by an in-plane anisotropy of the strain, $\rho/\Delta_{lh}$. The presence of this simple VBM term mixes hole-Cr$^+$ states with different S$_z$ and allows optical transitions with a change of the Cr$^+$ spin. Such transitions appear as low intensity PL lines in the calculated spectra presented in a log scale in Fig.~\ref{FigMap}(b). These lines are shifted from the main transitions ({\it i.e.} spin conserving transitions) by an energy which depends on the magnetic anisotropy D$_0$. For instance, the line pointed by an arrow in Fig.~\ref{FigMap}(b) corresponds to a transition with a change of the Cr$^+$ spin from $\pm 3/2$ to $\pm 5/2$ respectively. It is shifted from the spin conserving line by 4$D_0$. Such transitions appear sometimes in the PL spectra of dots with a large central gap (see QD1) but are usually hidden in the pedestal of the broadened spin conserving PL lines. 

To explain details of the PL spectra under magnetic field and in particular characteristic anti-crossings on the low energy line one has to include the presence of VBM induced by shear strain. This term is in particular responsible for anti-crossing $x1$ in Fig.~\ref{FigMap}(c). A negative value of D$_0$ is needed to control the position of $x1$. This negative value of D$_0$ associated with a fine structure term $E$ are also responsible for anti-crossing $x2$ in Fig.~\ref{FigMap}(c) and to some additional anti-crossings at low magnetic field on the high energy lines in $\sigma-$ polarization.

\section{Resonant photo-luminescence of X$^+$-Cr$^+$.}

The multi-level spin structure in a positively charged Cr$^+$-doped QD, including possible perturbation terms which do not commute with $D_0S_z^2$ or J$_z$S$_z$, results in a complex spin dynamics. A faster change of the spin state for X$^+$-Cr$^+$ compared to the spin evolution time in the hole-Cr$^+$ complex can be a source of optical pumping under resonant excitation. Resonant optical pumping was recently observed for an excitation on the high energy line of X$^+$-Cr$^+$ which addresses the low energy states of the hole-Cr$^+$ nano-magnet with M$_z=\pm$4 (see the energy level scheme in Fig.\ref{FigMap} and ref. \cite{Tiwari2021}).

A change of the spin in a Cr$^+$-doped QD can be induced by possible non-diagonal terms in the X$^+$-Cr$^+$ or hole-Cr$^+$ Hamiltonians. It can also result from a combination of spin-flip of the electron (in a few $\mu s$ range \cite{Legall2012}), spin-flip of the hole, independent spin-flip of the Cr$^+$ induced by the interaction with phonons and possible hole-Cr$^+$ flip-flops \cite{Lafuente2018}. In addition, it was already observed that the spin of a Cr atom can be sensitive to optically generated non-equilibrium phonons \cite{Tiwari2020,Tiwari2021}. As the spin splitting are significantly different in the ground state (overall splitting of the hole-Cr$^+$ states given by $5\times 3/2 \times I_{hCr^+}$) and in the excited state (X$^+$-Cr$^+$ levels dominated by the Cr$^+$ fine structure and split by $6\times D_0$), the spin dynamics in each complex is also expected to strongly differ. 

In a magnetic QD, the intensity distribution of the resonant PL can be used to reveal the main spin-flip channels \cite{Lafuente2015}. Under resonant excitation of X$^+$-Cr$^+$ and in a steady state regime, the amplitude of the PL results from (i) the efficiency of absorption, (ii) the population of the initial hole-Cr$^+$ state resonantly excited which is eventually decreased by a pumping effect and (iii) the efficiency of the spin transfer among the spin levels during the lifetime of the charged exciton.

\subsection{Resonant PL in the presence of non-resonant excitation}

\begin{figure}[hbt]
\centering
\includegraphics[width=1\linewidth]{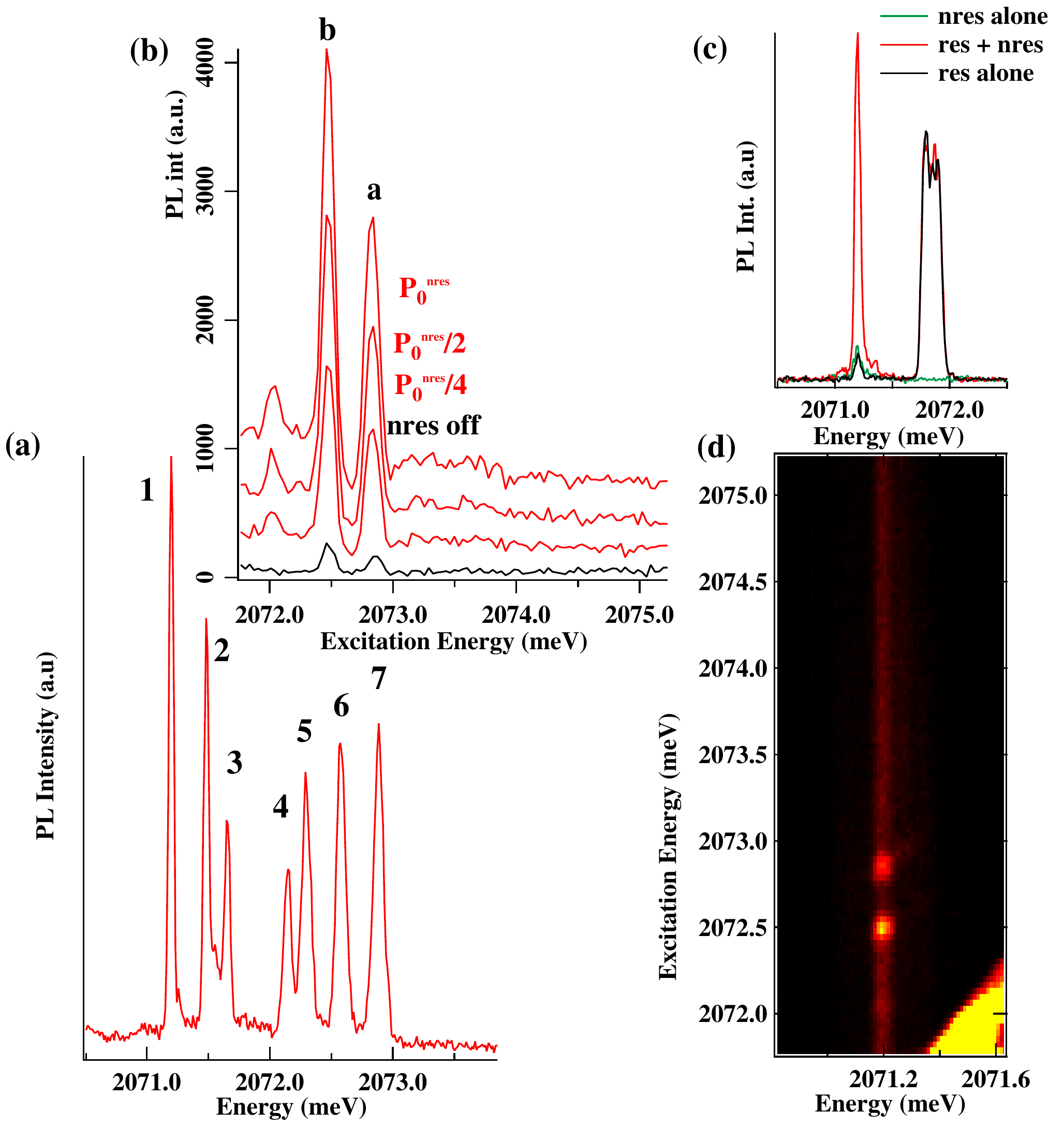}
\caption{Resonant PL detected on the low energy line 1 of X$^+$-Cr$^+$ in QD2 obtained in cross-linear excitation/detection configuration. The resonant PL is enhanced in the presence of an additional non-resonant (nres) excitation (at 568 nm). (a) PL spectra of X$^+$-Cr$^+$ in QD2. (b) PL excitation spectra obtained for variable non-resonant excitation power. (c) Resonant PL spectra obtained for an excitation on resonance $b$. (d) Intensity map of the PL excitation spectra.}
\label{Fig4}
\end{figure}

The main feature of the resonant PL of X$^+$-Cr$^+$ of QD2 is presented in Fig.~\ref{Fig4}. In these experiments the PL is detected on the low energy line $1$ for an excitation scanned on the high energy side of the spectra in a cross-linear excitation/detection configuration ($\pi_{cross}$).  

At high resonant excitation power, resonances appear around the two high energy lines $6$ and $7$ (black curve in figure Fig.~\ref{Fig4}(b)) when the PL is detected on line $1$. The resonant PL is enhanced when an additional non-resonant laser tuned on the energy range of the excited states of the QD, below the ZnTe barrier, is added. This non-resonant laser, at 568 nm in the experiment presented in Fig. \ref{Fig4}, produces a weak PL but has an influence on the intensity of the resonant PL. The amplitude of the resonant PL is enhanced by the increase of the power of the non-resonant excitation and a weak absorption peak appears around the energy of line $4$ (Fig.~\ref{Fig4}(b)).

Resonances at energies corresponding to the two high energy lines $6$ and $7$ are clearly observed whereas the contribution of an absorption on lines $4$ and $5$ is absent or very weak (Fig.~\ref{Fig4}(d)). Absorption $b$ appears to be slightly shifted from line $6$. Whereas the splitting between PL lines $6$ and $7$ is around 300 $\mu$eV, resonances $a$ and $b$ in the PL excitation spectra are split by about 390 $\mu$eV.

\subsection{Energy distribution and polarization of the resonant PL}

\begin{figure}[h!]
\centering
\includegraphics[width=1.0\linewidth]{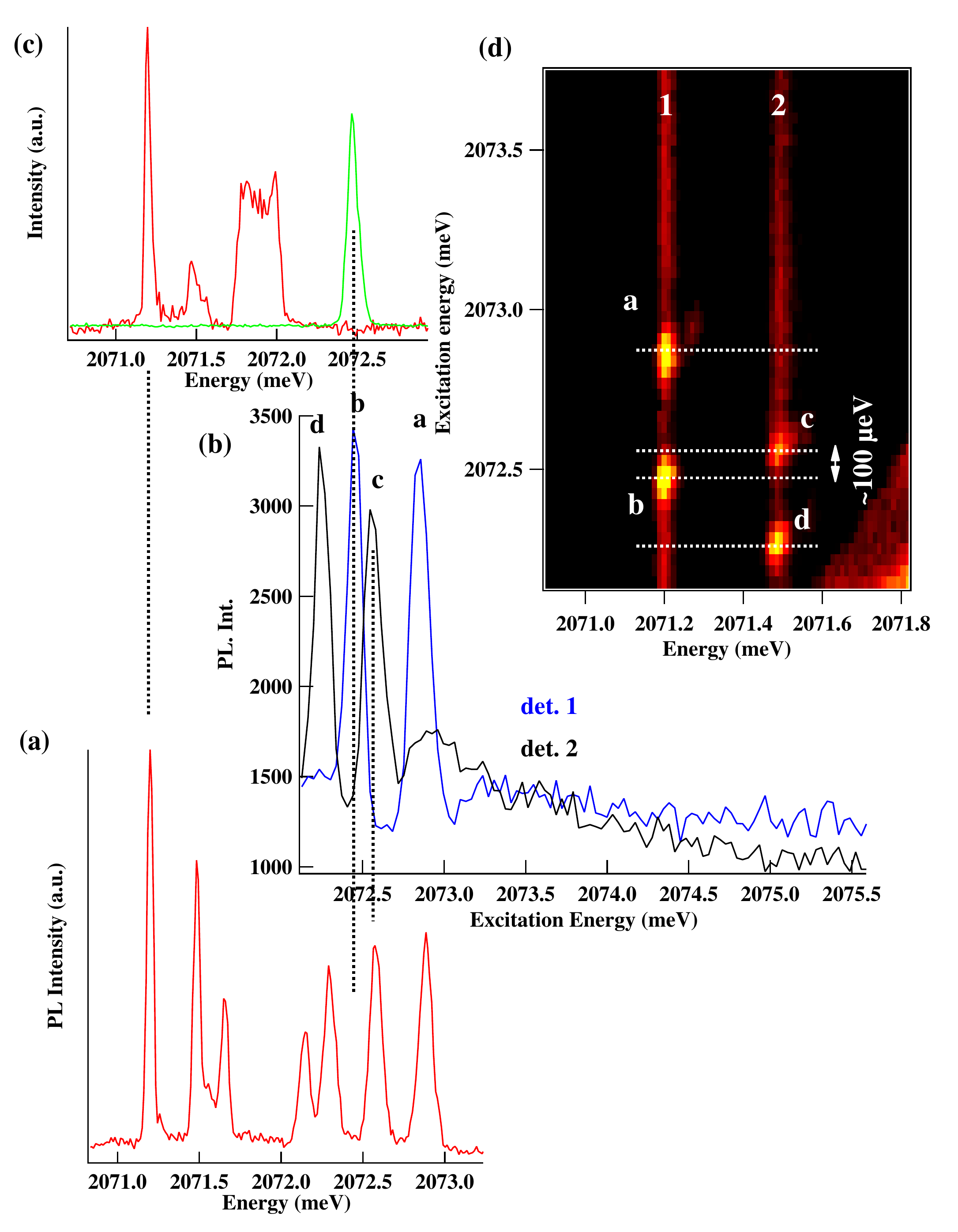}
\caption{PL of the two low energy lines 1 and 2 of  X$^+$-Cr$^+$ in QD2 for an excitation scanned on the high energy side of the spectra in $\pi_{cross}$ configuration. All the measurements are performed under an additional weak non-resonant excitation at 568 nm. (a) Reference non-resonant PL. (b) PL intensity as a function of the excitation energy for a detection on lines 1 (det. 1)  and 2 (det. 2). (c) Resonant PL for an excitation on $b$ together with the corresponding excitation laser line (green). (d) PL excitation intensity map.}
\label{FigPLE}
\end{figure}

A simultaneous observation of lines $1$ and $2$ when the resonant laser is scanned on the high energy side of the spectra permits to reveal the relative position of the absorption lines and obtain more information on the relaxation processes among X$^+$-Cr$^+$. The corresponding excitation spectra obtained for a fixed non-resonant excitation power and in $\pi_{cross}$ configuration are presented in Fig.~\ref{FigPLE}. The simultaneous observation of the two low energy lines permits to rule out some possible uncertainty on the excitation laser energy and to clearly measure the relative energy position of the different absorption lines.

The main results of these resonant PL experiments are the following: (i) An excitation on line $7$ produces a PL on line $1$ (resonance $a$ in Fig.~\ref{FigPLE}) and no resonant PL is observed on line $2$. (ii) An excitation on line $6$ produces a PL on line $2$ (resonance $c$ in Fig.~\ref{FigPLE}) and no resonant PL is observed on line $1$. (iii) A PL on line $1$ is observed for an excitation slightly shifted on the low energy side of line $6$ (resonance $b$ in Fig.~\ref{FigPLE}). At this excitation energy, no resonant PL is observed on line $2$. (iv) A PL is observed on line $2$ for an excitation on the low energy side of line $5$ (resonance $d$ in Fig.~\ref{FigPLE}). As already discussed for the experiments presented in Fig.~\ref{Fig4}, no PL is observed on line $1$ for an excitation around line 5.

These data can be understood with the energy level structure of X$^+$-Cr$^+$ which is dominated by the magnetic anisotropy term D$_0$S$_z^2$ (see section III and reference \cite{Tiwari2021}). Consequently, among X$^+$-Cr$^+$, the spin states $S_z=\pm5/2$, $S_z=\pm3/2$ and $S_z=\pm1/2$ are respectively degenerated (see Fig.~\ref{FigMap}(a)).

The observation of PL only on line $1$ (Cr$^+$ spin state $S_z=\pm$5/2) for an excitation on line $7$ (Cr$^+$ spin state $S_z=\pm$5/2) and the observation of PL only on line $2$ (Cr$^+$ spin state $S_z=\pm$3/2) for an excitation on line $6$ (Cr$^+$ spin state $S_z=\pm$3/2) show that under resonant excitation $\vert S_z\vert$ is well conserved during the lifetime of X$^+$ (around 250 $ps$). In particular, there is no detectable transfer from $\pm 3/2$ to the lowest energy states $\pm 5/2$, a spin flip channel that would involve an emission of low energy acoustic phonons.

\begin{figure}[hbt]
\centering
\includegraphics[width=1.0\linewidth]{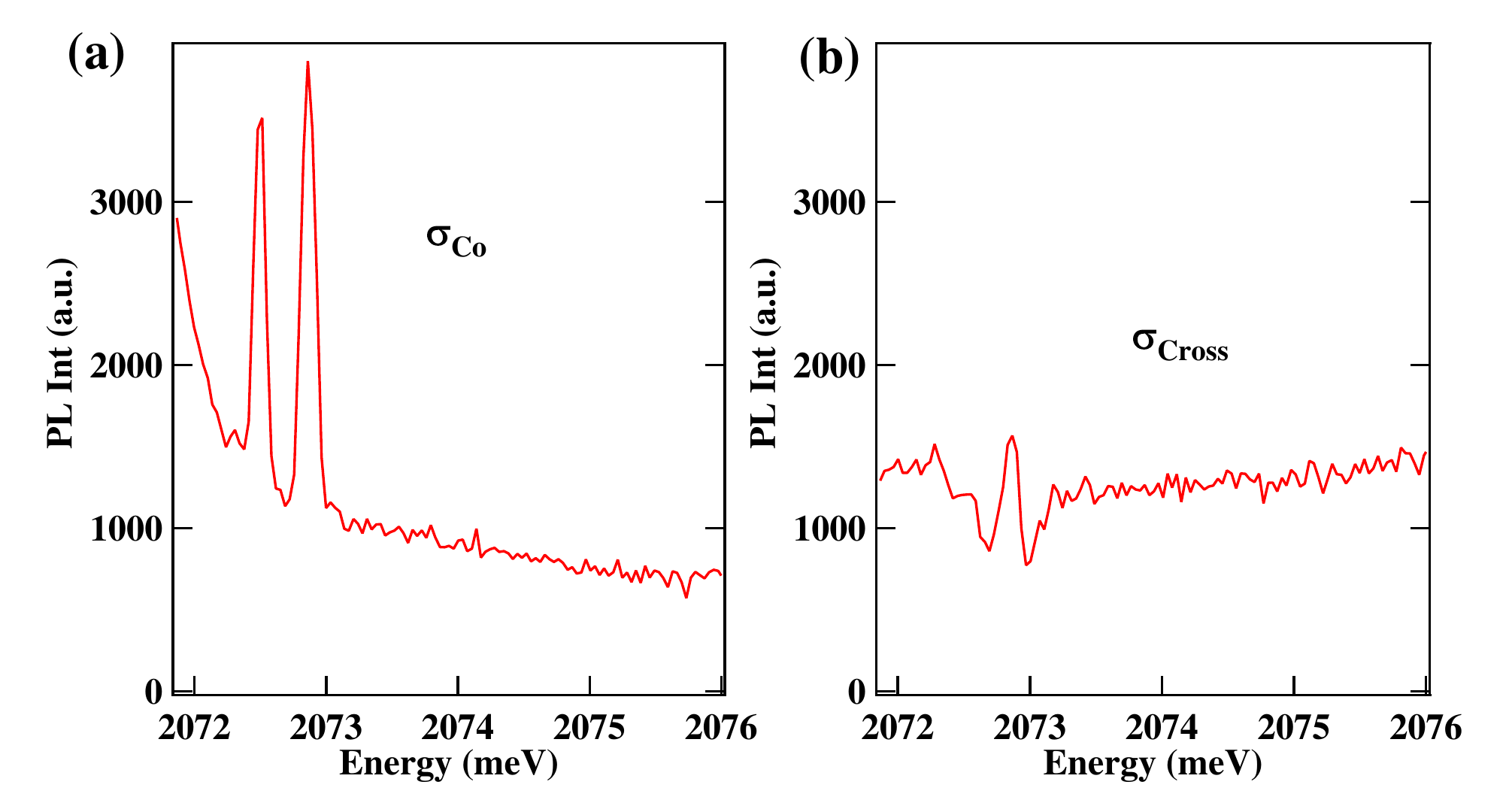}
\caption{PL excitation spectra detected on line $1$ of X$^+$-Cr$^+$ obtained in $\sigma_{co}$ (a) and $\sigma_{cross}$ (b) excitation/detection configuration and with a fixed non-resonant (at 568 nm) excitation power. Let us note that the background level of the non-resonant PL is slightly larger in $\sigma_{cross}$ than in $\sigma_{co}$. Because of the configuration of the set-up the $\sigma_{co}$ polarization for the resonant excitation corresponds to crossed circular polarization for the non-resonant excitation. For an excitation on the excited states range of the dots the non-resonant PL is slightly co-polarized.}
\label{Fig4bis}
\end{figure}

Additional information on the spin dynamics of X$^+$-Cr$^+$ can be obtained with circularly polarized resonant excitation. Resonant PL detected on line $1$ for $\sigma_{co}$ and $\sigma_{cross}$ configurations are presented in Fig.~\ref{Fig4bis}. As in the case of linearly polarized excitation, an increase of the resonant PL is observed when a high energy non-resonant excitation is added (see Fig.~\ref{Fig6}). The resonant PL is superposed to the constant background of non-resonant PL.

Clear absorption resonances appear in $\sigma_{co}$ for an excitation around the two high energy lines $6$ and $7$. The obtained resonant PL is mainly co-polarized with the excitation and only a weak cross-circularly polarized contribution is observed for an excitation on the high energy line $7$. At low excitation power, this cross-circular resonant PL appears at the center of a narrow hollow in the non-resonant PL background. Such contribution in cross-circular polarization is not observed for an excitation on the other X$^+$-Cr$^+$ lines. In particular, no $\sigma_{cross}$ signal is observed for an excitation around line $6$ where a clear $\sigma_{co}$ contribution is however obtained.

\subsection{Discussion on the origin of the resonant PL.}

A $\sigma_{cross}$ emission on the low energy line $1$ for an excitation on the high energy line $7$ corresponds to spin-flip of the electron with a conservation of the Cr$^+$ spin. The observed very weak $\sigma_{cross}$ component shows that a spin flip of the electron is unlikely. Electron-Cr$^+$ flip-flops are also unlikely. They would result in a dominant $\sigma_{cross}$ resonant PL with a transfer of the Cr$^+$ spin, from +5/2 to +3/2 for instance. This is the situation observed in positively charged Mn$^{2+}$-doped QDs where, in striking contrast with the Cr$^+$ case, the fine structure term D$_0$ is weak, the electron-Mn$^{2+}$ exchange interaction is much larger and dominates the structure of the energy levels of X$^+$-Mn$^{2+}$ \cite{Lafuente2015}. 

The transfer of excitation with a conservation of the circular polarization (dominant signal in $\sigma_{co}$) between lines $7$ and $1$ (between lines $6$ and $2$) corresponds to a transfer among $\pm5/2$ ($\pm3/2$) with a conservation of the electron spin. It shows that spin-flips between $\pm S_z$ can occur and dominate over the spin-flip of the electron and the interaction with low energy acoustic phonons that would induce a transfer from $\pm 3/2$ to the low energy states $\pm 5/2$. Such spin-flips among $\pm S_z$ correspond to a so-called spin-tunneling. Spin tunneling is well known to significantly influence the spin dynamics in weakly split molecular nano-magnet \cite{nanomagnet,Wernsdorfer2008}. These tunneling processes can be faster than the interaction with low energy acoustic phonons. In a single phonon process, the spin-flip probability evolves as the power of five of the energy splitting between the considered spin levels \cite{Cao2011,Tsit2005,Chud2005}. Within the weakly split X$^+$-Cr$^+$ system, the interaction with phonons is consequently reduced and no significant Cr$^+$ spin-flip between $\pm$5/2 and $\pm$3/2 and between $\pm$3/2 and $\pm$1/2 occur during the lifetime of X$^+$. 

This process explains resonances $a$ and $c$ in the PL excitation spectra. For resonance $a$, a laser resonant on line $7$ is absorbed when the Cr$^+$ spin is $\pm 5/2$ and creates an $X^+$-Cr$^+$ in its lowest energy state. It can recombine at the same energy with a conservation of the Cr$^+$ spin ({\it i.e.} strictly resonant fluorescence) or after a transfer from +5/2 to -5/2 (or from -5/2 to +5/2).  The second process gives some PL in $\sigma_{co}$ on the low energy line 1. In the case of resonance $c$, a laser resonant on line $6$ is absorbed when the Cr$^+$ spin state is $\pm3/2$. X$^+$ can recombine at the same energy or on line $2$ with the same circular polarization after a transfer from +3/2 to -3/2 (or from -3/2 to +3/2). 

Such process can induce an optical pumping of the state under resonant excitation ({\it i.e.} empty the resonantly excited spin state, S$_z$=+5/2 in the case of an excitation on 7). This pumping reduces the absorption probability and leads to a decrease of the resonant PL intensity. The resonant PL can be restored by the high energy non-resonant excitation which fasten the hole-Cr$^+$ spin relaxation. As we will see in more detail later, this explains the dependence of the resonant PL on the power of the high energy ({\it i.e.} not spin selective) excitation. 

These experiments show that the X$^+$-Cr$^+$ system behaves like a nano-magnet with a magnetic anisotropy $D_0S_z^2$. The degenerate spin states $\pm1/2$, $\pm3/2$ and $\pm5/2$ respectively can be weakly mixed by perturbation terms that do not commute with S$_z$. Such coupling are at the origin of spin tunneling that dominates over the coupling with low energy acoustic phonons among X$^+$-Cr$^+$ and gives rise to the observed weak resonant PL. Similar behavior is often observed in molecular nano-magnets where spin tunneling dominates their spin dynamics \cite{nanomagnet,Wernsdorfer2008}.

Additional absorption resonances (resonances $b$ and $d$), slightly shifted from the emissions lines, appear in the PL excitation spectra of Fig.~\ref{FigPLE}. In addition to the main optical transitions (labeled from $1$ to $7$ in Fig.~\ref{Fig4}) which conserve the Cr$^+$ spin, the presence of Hamiltonian terms which do not commute with S$_z$ allows some optical transitions with a change of S$_z$ by one unit. Traces of these transitions appear at zero magnetic field in some of the dots (see for instance QD1 in Fig.~\ref{FigLevels}) but are more likely observed under magnetic field when they are mixed with optically allowed transitions (see magneto-optic maps in Fig.~\ref{FigLevels}). These transitions appear in the calculated spectra in the presence of VBM terms (Fig.~\ref{FigMap}(b)). Transitions which conserve the Cr$^+$ spin (dominant PL lines) and transitions which occur with a change of one of S$_z$ are shifted by an energy which depends on the magnetic anisotropy $D_0$.

For an excitation around line $6$ (S$_z$=$\pm$3/2), the optical transition can occur with a change of the Cr$^+$ spin by one unit starting from a hole-Cr$^+$ state with S$_z$=$\pm3/2$ and creating an X$^+$-Cr$^+$ state with S$_z$=$\pm5/2$. This transition, which corresponds to resonance $b$, is shifted from line $6$ by an energy $((5/2)^2-(3/2)^2)D_0$=$4D_0<0$. 

Let's consider for instance a $\sigma-$ excitation scanned around line $6$. It addresses the hole-Cr$^+$ state $\vert S_z=+3/2,\Uparrow_h\rangle$. This levels is degenerate with $\vert S_z=-3/2,\Downarrow_h\rangle$. The latter is coupled by the VBM with the state $\vert S_z=-5/2,\Uparrow_h\rangle$ (see green arrows in Fig.~\ref{FigMap}(a)). The $\sigma-$ excitation can then induce a transition towards the X$^+$-Cr$^+$ state $\vert S_z=-5/2,\Uparrow_h \Downarrow_h \uparrow_e\rangle$. This transition is shifted from the spin conserving one (from $\vert S_z=+3/2,\Uparrow_h\rangle$ to $\vert S_z=+3/2,\Uparrow_h \Downarrow_h \uparrow_e\rangle$) by $4D_0<0$. $\vert S_z=-5/2,\Uparrow_h \Downarrow_h \uparrow_e\rangle$ recombines on the low energy line $1$ by emitting a $\sigma-$ photon. This explains the weak $\sigma_{co}$ emission for an excitation on resonance $b$. This forbidden optical transition is a source of optical pumping for the states $\vert S_z=-3/2,\Downarrow_h\rangle$ which is excited at this energy. The observed energy shift permits a direct determination of the magnetic anisotropy of the Cr$^+$ atom and for QD2 $4 D_0 \approx -100 \mu eV$.

The same mechanism explains the $\sigma_{co}$ emission for an excitation on resonance $d$. Transition $5$ is associated with the Cr$^+$ spin states S$_z=\pm1/2$. An absorption on line $5$ can produce a PL on $5$ or on $3$ and $4$ (mixed levels) after a transfer which conserves $\vert S_z\vert$. As in the case of resonance $b$, an excitation close to line $5$ can also induce a transition from S$_z$=$\pm1/2$ to S$_z$=$\pm3/2$. Neglecting the VBM which can have an influence on the energy of these levels close to the central gap, this transition is shifted by $((3/2)^2-(1/2)^2)D_0$=$2D_0<0$ from line $5$. The created X$^+$-Cr$^+$ associated with a Cr$^+$ spin S$_z=\pm$3/2 recombines on line $2$, as observed for resonance $d$. As in the case of resonance $b$, this transition can induce an optical pumping of the resonantly excited state. 

\begin{figure}[h!]
\centering
\includegraphics[width=1.0\linewidth]{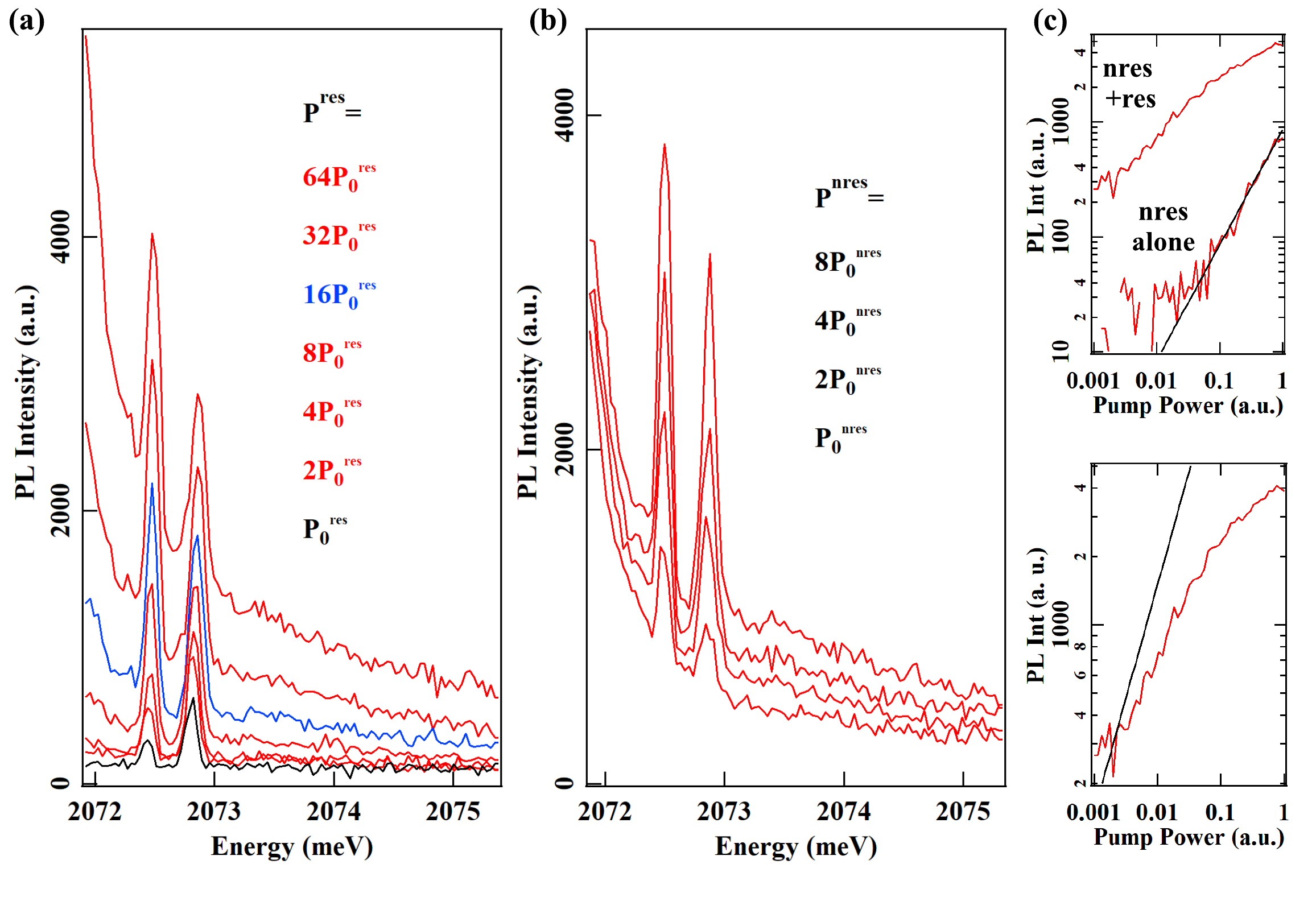}
\caption{Power dependence the PL excitation spectra detected on the low energy line $1$ in $\sigma_{co}$ configuration. (a) Dependence on the power of the resonant excitation for a fixed non-resonant excitation power. (b) Dependence on the power of the non-resonant excitation for a fixed resonant excitation. (c) Evolution of the PL intensity as a function of the non-resonant excitation power on line 7. In the bottom panel the contribution of the non-resonant PL is subtracted from the total PL. Black lines are linear fits.}
\label{Fig6}
\end{figure}

Resonances $a$ and $c$ on one side and resonances $b$ and $d$ on the other side have a different origin. Resonances $b$ and $d$ are in a first approximation forbidden optical transitions with a change of S$_z$ and are expected to have a weak oscillator strength. The different nature of these transitions is confirmed by their excitation power dependence. The intensity of the circularly polarized resonant PL depends both on the resonant and non-resonant excitation power. This is presented in Fig.~\ref{Fig6} where the PL excitation spectra detected on line $1$ are displayed for varying resonant (Fig.~\ref{Fig6} (a)) or non-resonant (Fig.~\ref{Fig6} (b)) excitation intensity in the $\sigma_{co}$ configuration.

For a fixed non-resonant excitation, with the increase of the resonant power a contribution of a direct absorption in the high energy acoustic phonon side-band of the low energy line appears. At low resonant excitation power, the resonant PL obtained for an excitation on line $7$ is much larger than for an excitation around line $6$. At high excitation power their intensity is roughly equal. This confirms that these two transitions have a significantly different oscillator strength. The transition around line $6$ which has the weaker oscillator strength saturates at a higher excitation power than the optically allowed transition $7$ which has the largest intensity at low excitation power.

\section{Spin resonant optical pumping and spin heating.}

For a fixed resonant excitation intensity, increasing the non-resonant excitation increases the amplitude of each absorption peaks (Fig.~\ref{Fig6}(b)). For an excitation on the high energy line $7$, the intensity of the resonant co-circularly polarized PL is continuously tuned by the non-resonant excitation. Its evolution with the increase of the non-resonant excitation is sub-linear and a saturation is observed at high excitation (Fig.~\ref{Fig6}(c)). 

The dependence of the resonant PL on the intensity of the non-resonant excitation can be explained by the presence of the resonant optical pumping of the hole-Cr$^+$ spin \cite{Govorov2005}. The out of equilibrium population distribution created by the resonant pumping can be partially destroyed by the high energy excitation. The amplitude of the resonant PL is therefore restored. We will show in the following that the resonant optical pumping and its destruction during the non-resonant excitation can be observed in the time evolution of the resonant PL.

\subsection{Resonant PL under modulated excitation.}

\begin{figure}[h!]
\centering
\includegraphics[width=1.0\linewidth]{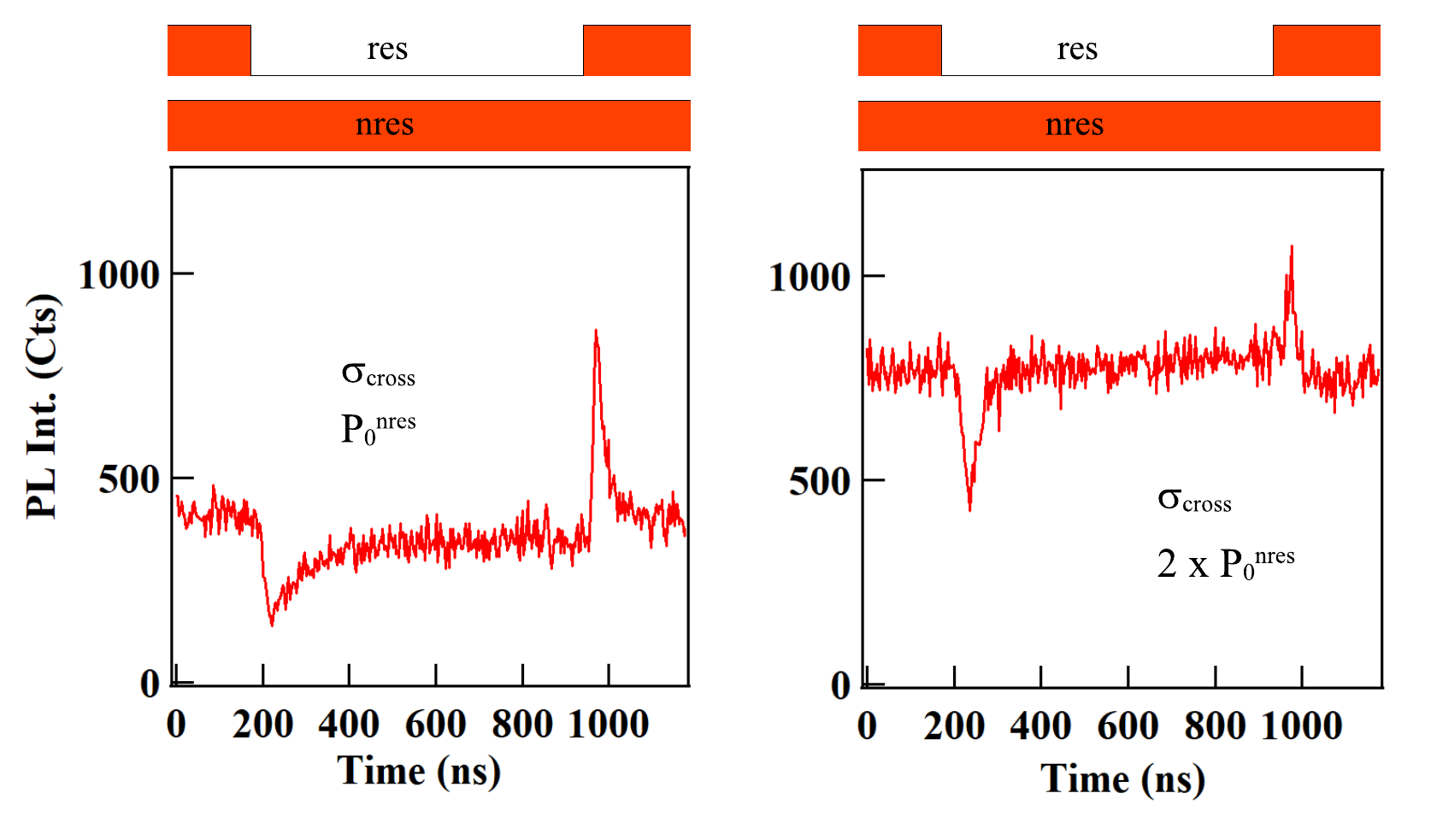}
\caption{Time resolved resonant PL detected on line $1$ of QD2 under a CW non-resonant excitation (568 nm) and a pulsed resonant excitation on line $7$. The transients obtained in $\sigma_{cross}$ configuration for two different non-resonant excitation powers are compared.}
\label{Fig15S}
\end{figure}

When a continuous wave (CW) non-resonant excitation is combined with a pulsed resonant excitation on the high energy line $7$, transients are observed in the resonant PL intensity of the low energy line $1$. This is presented in  Fig.~\ref{Fig15S} for a $\sigma_{cross}$ excitation/detection configuration. Despite the weak resonant signal (see Fig.~\ref{Fig4bis}), this excitation/detection configuration is of particular interest. In this configuration one excites and detects the same spin state of the Cr$^+$ (see the energy level scheme in Fig.~\ref{FigMap}). For instance, an excitation on line $7$ in $\sigma-$ polarization empties the Cr$^+$ spin state S$_z$=+5/2 in the hole-Cr$^+$ complex. The PL of line $1$ recorded in $\sigma+$ polarization is also proportional to the probability of occupation of the Cr$^+$ spin state S$_z$=+5/2.

The transient observed at the beginning of the resonant pulse corresponds to the optical pumping of the resonantly exited hole-Cr$^+$ spin state. The out-of-equilibrium spin occupation probability created by spin pumping is destroyed by the non-resonant excitation: This is directly observed in the CW signal just after the end of the resonant pulse. This appears as a progressive increase of the intensity of the non-resonant PL ({\it i.e.} filling of the spin state S$_z$=+5/2 after a $\sigma-$ resonant excitation on the high energy line). This heating of the hole-Cr$^+$ spin by the high energy excitation typically takes place in a few tens of $ns$. The speed of the spin heating increases with the increase of the non-resonant excitation power (Fig.~\ref{Fig15S}(b)). 

It is clear from these experiments that the resonant excitation affects the intensity of the non-resonant PL. This confirms the presence of the resonant pumping and the observation of a transient in the CW non-resonant PL reflects its destruction by the non-resonant excitation. These pumping experiments also show that under optical excitation spin fluctuations in the tens of $ns$ range takes place within the X$^+$-Cr$^+$ or the hole-Cr$^+$ complexes.

\subsection{Spin fluctuations observed in the auto-correlation of the PL intensity.}

Spin fluctuations within a positively charged Cr$^+$-doped QD directly influence the statistics of the photons emitted by individual lines of X$^+$-Cr$^+$. This statistics can be observed in the auto-correlation of the PL intensity. To estimate the auto-correlation signal we used a Hanbury Brown-Twiss set-up with a time resolution of about 700 ns \cite{Lafuente2016APL}. In this experiment, the detection at t=0 of a photon at a given energy and polarization indicates that the hole-Cr$^+$ system is in a particular spin state. The delayed detection of a second photon with the same energy and polarization depends on the probability of conservation of the spin state. This photons' coincidence measurement is a good approximation of the auto-correlation of the PL intensity as far as the considered time delay remains weaker than the inverse of the photon collection rate limited in these experiments to a few kHz.

\begin{figure}[h!]
\centering
\includegraphics[width=1.0\linewidth]{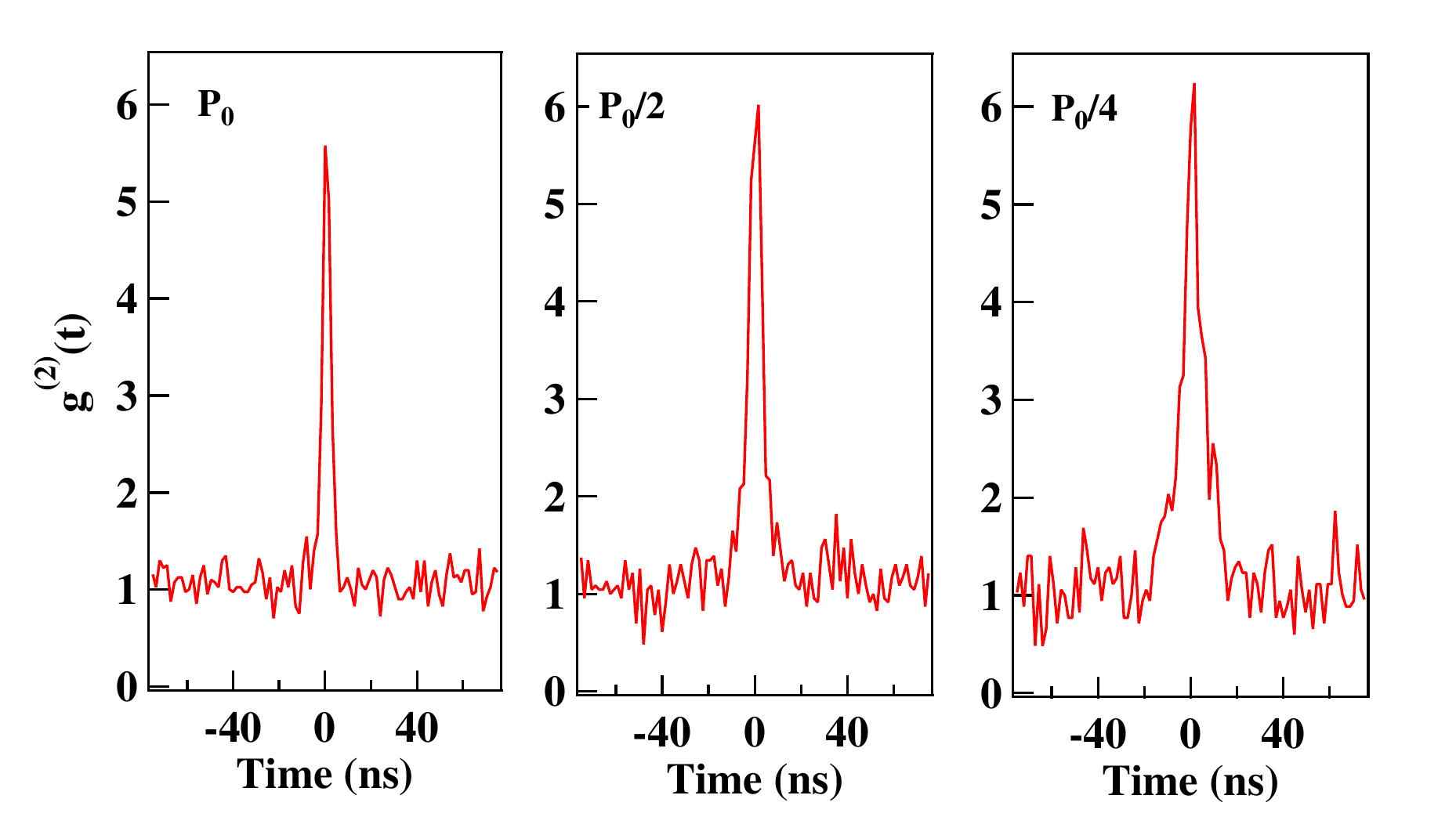}
\caption{Auto-correlation of the PL intensity of the low energy line of X$^+$-Cr$^+$ in QD2 for three different non-resonant ($\lambda$= 578.5 nm) excitation powers.}
\label{Fig14S}
\end{figure}

Figure \ref{Fig14S} presents the auto-correlation of the PL intensity of line $1$ of QD2 for an excitation at 578.5 nm on excited states of the dot. Because of the limited time resolution of the setup and the short lifetime of X$^+$ (around 250 ps), the antibunching characteristic of a single photon emitter is not observed. However, a large photon bunching is obtained with a characteristic width of a few $ns$. This bunching is a signature of fluctuations in the PL intensity. The width of the bunching reflects the speed of the spin-flips whereas its amplitude depends on the number of available states for the spin relaxation. The speed of the spin fluctuations increases with the increase of the excitation power. 

PL intensity fluctuations can arise both from the hole-Cr$^+$ and the X$^+$-Cr$^+$ spin dynamics. As coincidence events in auto-correlation measurements depend on the square of the photon count rate, these experiments are performed under high excitation intensity where the probability of presence of X$^+$ in the dot is large. 

The increase of spin fluctuations with the non-resonant excitation power can have two origins. First the spin dynamics is faster within the X$^+$-Cr$^+$ than for the hole-Cr$^+$ complex and spin-flips increase with the increase of the probability of presence of X$^+$. Secondly, the high energy and high intensity optical excitation generates non-equilibrium acoustic phonons that can induce spin-flips of the Cr$^+$, in particular within the split hole-Cr$^+$ complex \cite{Tiwari2021, Tiwari2020}. The two processes probably contribute to the acceleration of the spin dynamics. This experiment does not permit to identify the dominant contribution. However it confirms that (i) under excitation spin fluctuations in the $ns$ range take place and (ii) that the speed of these fluctuations is very sensitive to the excitation power.

\subsection{Analysis of the resonant pumping transient.}

\begin{figure}[h!]
\centering
\includegraphics[width=1.0\linewidth]{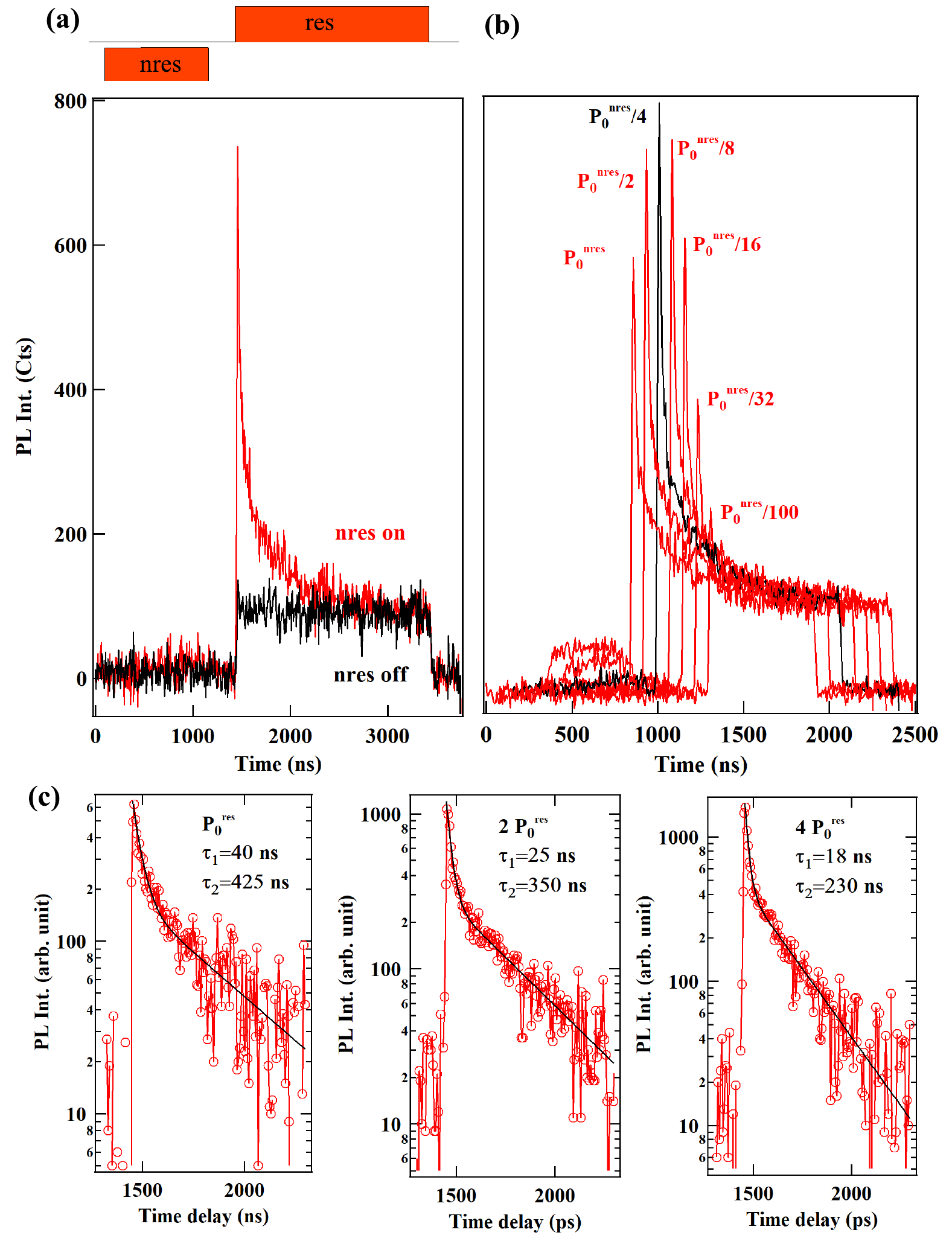}
\caption{(a) Optical spin pumping with separated non-resonant and resonant pulses for a resonant excitation on line $7$ and a co-circular detection on line $1$. (b) Evolution of the resonant pumping transients as a function of the non-resonant excitation power. Traces are shifted along the time axis for clarity. (c) Evolution of the resonant pumping transients as a function of the resonant excitation power. Black lines are bi-exponential fits with characteristics times $\tau_1$ and $\tau_2$.}
\label{Figpulsepump}
\end{figure}

To analyze the dynamics of the resonant pumping, the non-resonant and the resonant excitation can be both pulsed and separated in time (Fig.~\ref{Figpulsepump} (a)). For an excitation on line $7$, the $\sigma_{co}$ PL of line $1$ can be used to monitor the dynamics of the resonant absorption. The $\sigma_{co}$ configuration permits to obtain a much larger resonant signal. This was already observed in the steady state regime under CW resonant excitation (Fig.~\ref{Fig4bis}) and is also well pronounced in the pumping transients \cite{Tiwari2021}. Resonant pumping transients for different non-resonant and resonant excitation powers are presented in Fig.~\ref{Figpulsepump}(b) and Fig.~\ref{Figpulsepump}(c) respectively.

With the increase of the power of the non-resonant excitation and for a given resonant excitation (Fig.~\ref{Figpulsepump}(b)), the amplitude of the resonant pumping transient first increases, reaches a maximum and then decreases. An influence of the non-resonant excitation on the resonant pumping transient is observed even if there is not a significant non-resonant PL. Such behavior was already observed in the case of QDs containing a Cr$^{2+}$ ion \cite{Lafuente2018}. It was interpreted as the influence of locally generated acoustic phonons on the Cr$^{2+}$ spin split by a large strain induced magnetic anisotropy. This is also the case for the hole-Cr$^+$ system split by the ferromagnetic exchange interaction of the hole with the magnetic atom. At low excitation power, the non-resonant excitation pulse partially destroys the optical pumping ({\it i.e} populates the low energy state emptied by the resonant pumping). This produces an increase of the amplitude of the pumping transient. However, at high non-resonant excitation intensity, the effective spin temperature of the hole-Cr$^+$ system at the end of the non-resonant pulse increases and become much larger than the lattice temperature. This reduces the occupation probability of the hole-Cr$^+$ ground states and consequently the initial amplitude of the resonant PL transient. As we will discuss later, the effective temperature of the hole-Cr$^+$ spin, much larger than the temperature of the lattice, is induced by the interaction with the non-equilibrium acoustic phonons generated by the optical excitation. 

A careful analysis of the pumping signal also reveals that the transients are not simply mono-exponential (Fig.~\ref{Figpulsepump}(c)). They can be described by a bi-exponential decay whose characteristics times depend on the resonant excitation power. The dominant fast initial decay which takes place in a few tens of $ns$ is followed by a second transient with a characteristic time of a few hundreds $ns$. Both transients depend on the resonant excitation power.

The first decay corresponds to the decrease of the population of the state which is resonantly excited. The presence of the second decay shows that the state which is excited is powered by a transfer from other hole-Cr$^+$ levels. These levels, linked to the ground state, are also partially emptied at the end of the pumping sequence. Such transfer can be induced by all the non-diagonal terms of the hole-Cr$^+$ Hamiltonian affecting the ground states. 

The two ground states, with parallel hole and Cr spins, are in particular linked to the other levels by the $E$ term which couples Cr$^+$ spin states separated by two units. At the end of the pumping sequence the resonantly excited ground state and the coupled states are partially emptied. The transfer time depends on the strength of the coupling and on dephasing processes. Because of the weaker splitting, couplings induced by $E$ (or any other non-diagonal term) are much more efficient within X$^+$-Cr$^+$ than within the hole-Cr$^+$ system. Spin flips of the Cr$^+$ will consequently be much slower in the ground state than in the excited state allowing the resonant optical spin pumping. 

These detailed analyses of the pumping transients confirms the complex dynamics of this multi-levels spin system that would deserve to be modeled in detail. We can nevertheless conclude from these time resolved pumping experiments that the dynamics in this Cr$^+$ system strongly depends on the condition of optical excitation, both resonant and non-resonant.

\subsection{Effective Cr$^+$ spin temperature under non-resonant optical excitation}

\begin{figure}[h!]
\centering
\includegraphics[width=1.0\linewidth]{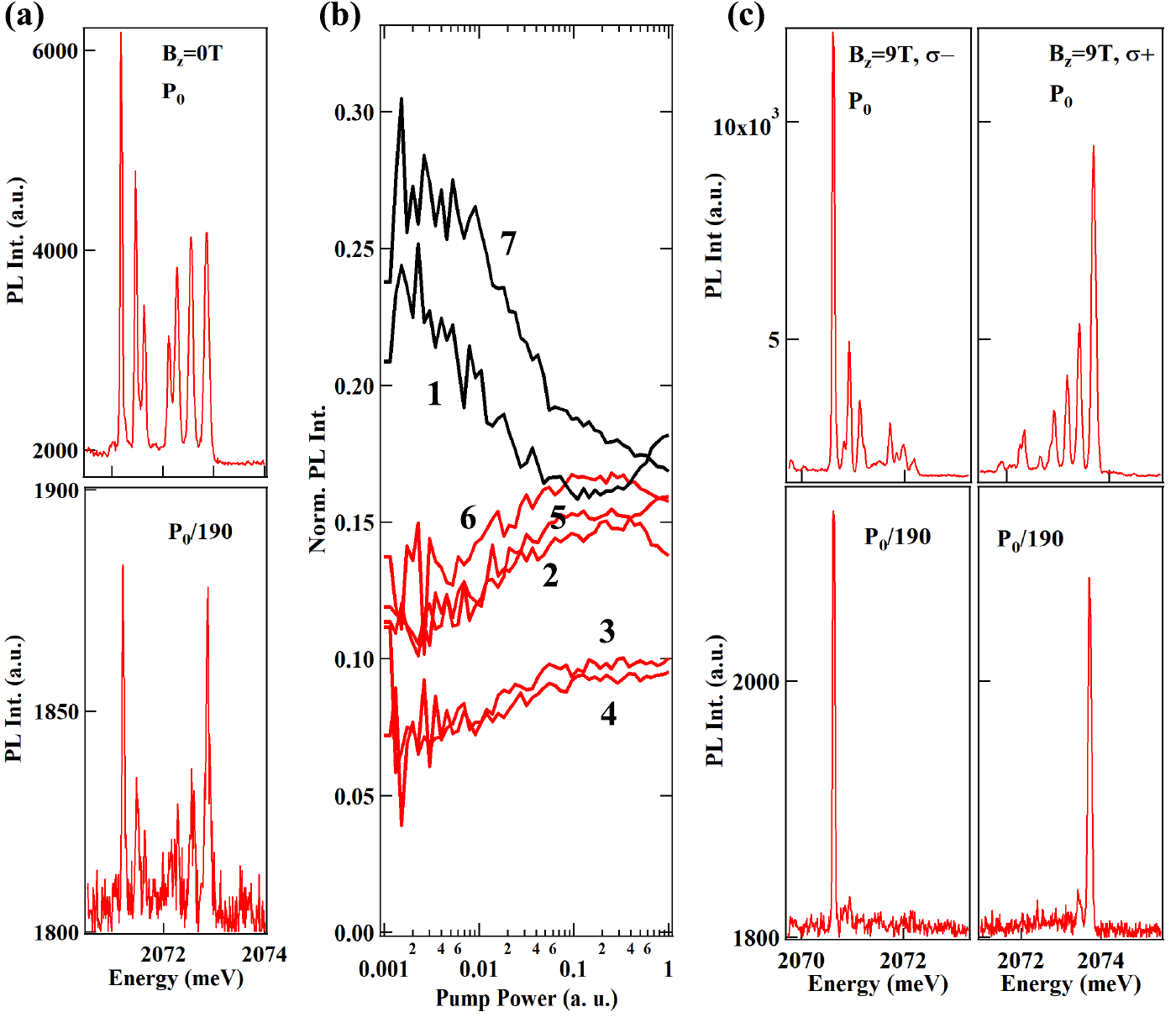}
\caption{Influence of the excitation power on the non-resonant PL intensity distribution observed at zero magnetic field (a) and under a longitudinal magnetic field B$_z$=9T in two circular polarization (c). (b) Power dependence of the PL intensity distribution of X$^+$-Cr$^+$ at zero magnetic field. The intensities of the two outer lines 1 and 7 are displayed in black.}
\label{Fig13S}
\end{figure}

The influence of the non-resonant optical excitation on the effective spin temperature of the Cr$^+$ can be observed in the PL intensity distribution of X$^+$-Cr$^+$. This distribution significantly depends on the excitation power. This is illustrated in the power dependence presented in Fig.~\ref{Fig13S} realized at a bath temperature of T=4.2K for an excitation on an excited state of the dot. In this dot (QD2) presenting a large hole-Cr$^+$ exchange interaction, at zero magnetic field and at weak excitation power the PL is mainly concentrated on the two outer lines. These lines correspond to the spin states S$_z$=$\pm5/2$ of the Cr$^+$. With the increase of the excitation power, the relative contribution of the PL of the two outer lines decreases and at high power all the seven lines are clearly observed. They present a similar integrated PL with a normalized intensity around 1/6, except for lines 3 and 4 which result from a splitting of the same hole-Cr$^+$ level and have an intensity around 1/12. For a linearly polarized non-resonant excitation, an equal intensity for all the PL lines is expected for an identical probability of occupation of all the 6 Cr$^+$ spin states.

A similar excitation power dependence is observed in the PL intensity distribution under a large magnetic field. Under B$_z$=9T, the thermalization on the low energy line in $\sigma-$ polarization and on the high energy line in $\sigma+$ polarization, characteristic of the ferromagnetic hole-Cr$^+$ coupling \cite{Tiwari2021}, is significantly enhanced at low excitation intensity where a single line is observed in the circularly polarized PL spectra (Fig.~\ref{Fig13S}(c)). 

At 4.2 K and in the absence of optical excitation, the hole-Cr$^+$ spin thermalize on the two ground states with angular momentum M$_z$=$\pm$4. At very low excitation intensity the QD is empty most of the time. If the spin relaxation among the X$^+$-Cr$^+$ complex is slower than the lifetime of X$^+$ (around 250 ps) the PL intensity distribution reflects this thermalization in the ground state and a maximum of PL is expected on the two outer lines.

The modification of the effective spin temperature of magnetic atoms in a semiconductor can result from the effect of the optical injection of carriers and/or the interaction with optically generated phonons \cite{Yakovlev2006,Furdyna1988}. Let's first analyze the influence of carriers in the particular case of a QD containing a hole-Cr$^+$ complex. Under non-resonant excitation, efficient spin relaxation of the hole occurs during the injection of the exciton: bright excitons (with anti-parallel electron and hole spins) or dark excitons (with parallel electron and hole spins) are injected in the dot with a similar probability. They can both form with the resident hole a X$^+$ that recombines optically. 

For an initial hole-Cr$^+$ on states M$_z$=$\pm$4 and in the absence of spin-flip during the lifetime of X$^+$-Cr$^+$, an injected bright exciton produces a PL on the high energy line and leaves the hole-Cr$^+$ to its ground state. The electron of an injected dark exciton optically recombines with the resident hole and gives a PL on the low energy line. This leaves the hole-Cr$^+$ system on the highest energy hole-Cr$^+$ states with anti-parallel hole and Cr$^+$ spins (M$_z$=$\pm$1). Starting from this high energy state, the most efficient relaxation process is a spin-flip of the hole that will drive the hole-Cr$^+$ to its ground state without changing the spin of the Cr$^+$. Cr$^+$ spin-flip induced by the spin lattice coupling and involving emission of phonons can also take place and hole-Cr$^+$ flip-flops induced by the presence of VBM can occur and contribute to a change of the Cr$^+$ spin \cite{Lafuente2018}. With these processes, the non-resonant optical injection of carriers can be a source for the increase of the effective spin temperature of the hole-Cr$^+$ system. Spin-flips among X$^+$-Cr$^+$ can also take place. They can be induced by non-diagonal terms of the Hamiltonian that do not commute with D$_0$S$_z$ ($E$ term for example). At high excitation intensity, when the probability of occupation of the QD is large, their contribution to the relaxation of the Cr$^+$ spin could also become significant. 

However, the thermalization on the two outer lines is also particularly pronounced under a large magnetic field (see Fig.~\ref{Fig13S}(c) for B$_z$=9T). Under a magnetic field and in the absence of excitation (or at low excitation power), the Cr$^+$ spin thermalizes on the spin states S$_z$=-5/2 and the hole-Cr$^+$ complex relaxes towards the spin states associated with S$_z$=-5/2. This corresponds to the low energy line of X$^+$-Cr$^+$ in $\sigma-$ polarization and to the high energy line in $\sigma+$ polarization. These lines dominate the PL spectra at low excitation power. At high excitation power all the lines are clearly observed reflecting a significant increase of the effective temperature of the Cr$^+$ spin. The power dependence under a large magnetic field is similar to the one observed at zero field and the spin-flip channels involved in the spin heating process are likely identical. 

Under a large field and for a hole-Cr$^+$ on one of the two ground states with S$_z$=-5/2, the injection of a bright exciton creates X$^+$-Cr$^+$ complex in its ground state. It recombines optically without inducing any change of the Cr$^+$ spin. The injection of a dark exciton also creates the X$^+$-Cr$^+$ complex in its ground state. It recombines optically towards the low energy hole-Cr$^+$ level with a change of the spin of the resident hole but does not affect the spin of the Cr$^+$. At high field (for instance B$_z$=9T in Fig.~\ref{Fig13S}(c)) the Zeeman splitting of X$^+$-Cr$^+$ is much larger than possible non-diagonal terms of the X$^+$-Cr$^+$ Hamiltonian and these terms cannot significantly contribute to the spin dynamics among X$^+$-Cr$^+$. Under a large magnetic field, the non-resonant injection of carriers in the dot does not induce any particular spin-flip channel and another source of spin heating has to be taken into account. 

The increase of the effective spin temperature can come from a direct interaction of the Cr$^+$ spin with phonons optically generated under CW high energy optical excitation. Such influence of out-of-equilibrium phonons, well known in diluted magnetic semiconductors \cite{Yakovlev2006}, was already observed in the spin relaxation after a heating pulse in the case of Cr$^{2+}$ \cite{Tiwari2020} and hole-Cr$^+$ \cite{Tiwari2021}. Under high energy optical excitation, the relaxation of carriers generates a continuous flow of acoustic phonons through the inelastic relaxation process. In the presence of a high population of phonons at the energy of the spin transitions, direct spin relaxation processes in which a single phonon is emitted or absorbed during the transition can occur. When higher energy phonon modes are populated in the presence of non-equilibrium phonons it is also possible to absorb and emit two high energy phonons to perform the spin transition \cite{Abragam}. These are indirect processes that can be faster than the direct transition because of the larger density of states for phonons at high energy. Both mechanisms induce an increase of the spin-lattice coupling with the increase of the density of phonons and can significantly modify the occupation probability of the Cr$^+$ spin states. This explains that a faster Cr$^+$ spin relaxation is obtained under optical excitation than in the dark even for a weak non-resonant PL ({\it i.e.} weak injection of carriers inside the dot) and that an efficient increase of the spin effective temperature is observed under high intensity optical excitation even under a large magnetic field.

\section{Acoustic phonons and resonant spin pumping.}

\begin{figure}[h!]
\centering
\includegraphics[width=1.0\linewidth]{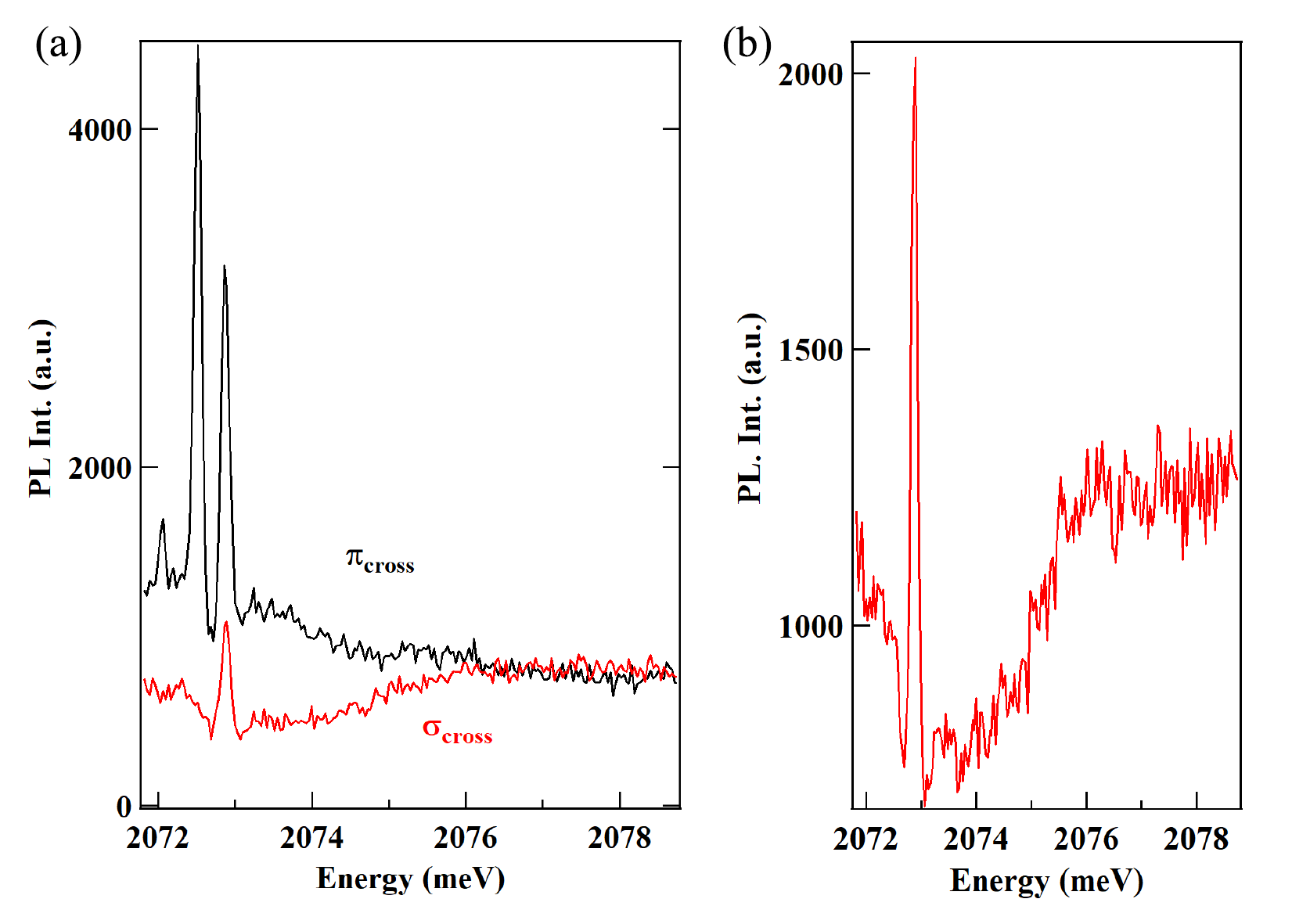}
\caption{Intensity of the PL of the low energy line $1$ for an excitation scanned around the high energy lines in cross-linear (black) and cross-circular (red) excitation/detection configurations. (b) Detail of the PL excitation spectra in $\sigma_{cross}$ configuration showing the contribution of the absorption in the acoustic phonon side-band around a resonant PL peak.}
\label{Fig8S}
\end{figure}

In addition to their complex multi-level spin structure, excitons in II-VI magnetic QDs strongly interact with the continuum of acoustic phonons \cite{Besombes2001}. The structure of the resonant PL can also be affected by this coupling. Details of the PL of line $1$ in QD2 for an excitation on the high energy side of X$^+$-Cr$^+$ in $\pi_{cross}$ and $\sigma_{cross}$ configurations are compared in Fig.~\ref{Fig8S}. In both measurements a fixed non-resonant excitation (at 568 nm, below the ZnTe barriers) is combined with the resonant laser. 

\begin{figure}[h!]
\centering
\includegraphics[width=1.0\linewidth]{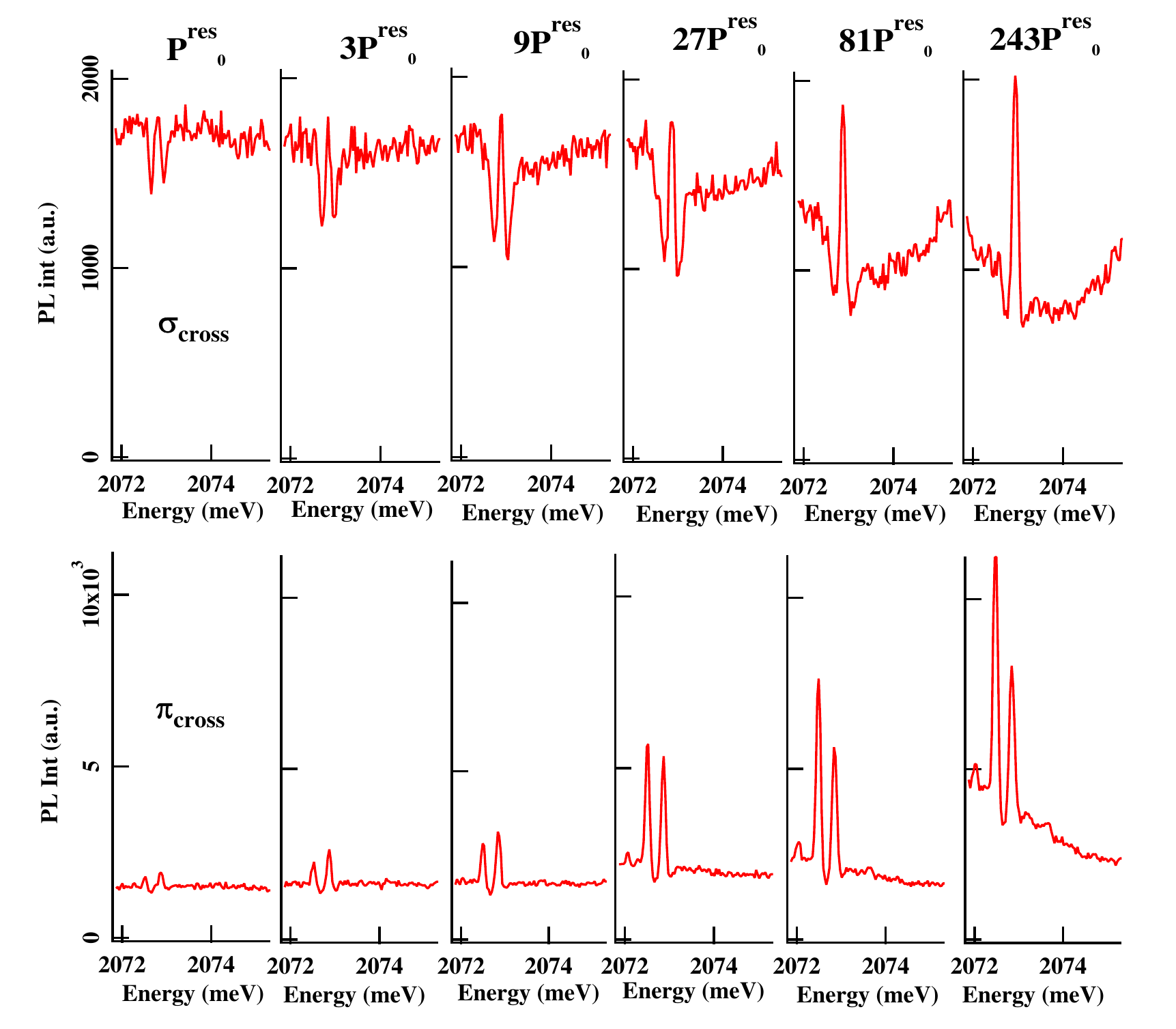}
\caption{Power dependence of the PL excitation spectra detected on the low energy line $1$ for $\sigma_{cross}$ (top panels) and $\pi_{cross}$ (bottom panels) configurations. All the PL excitation spectra are obtained with the same power of non-resonant excitation at 568 nm.}
\label{Fig8Sbis}
\end{figure}

In $\pi_{cross}$ configuration, in addition to the two resonances $a$ and $b$ around the two high energy lines, a background of absorption is also observed at high excitation power. The amplitude of the background increases with the reduction of the excitation energy and with the increase of the excitation power (Fig.\ref{Fig8Sbis}). This background of PL is due to a contribution of a direct absorption in the high energy acoustic phonon side-band of the low energy line $1$. Such background of absorption on the high energy side of a PL line can also be observed in non-magnetic QDs.

In $\sigma_{cross}$ configuration, when the resonant laser is scanned on the high energy side of the spectra, the PL is dominated by the weak contribution of the non-resonant excitation (background level). Because of the good conservation of the electron spin during the lifetime of X$^+$, the direct absorption in the acoustic phonon side-band of the low energy line is not observed ({\it i.e.} no increasing PL background at low excitation energy). In addition to the weak non-resonant PL, a resonant PL of line $1$ is observed for an excitation on line $7$. A similar signal is not observed for an excitation on the other lines and in particular around line $6$ which gives a large signal in $\pi_{cross}$ (or $\sigma_{co}$) configuration. This shows that the weak resonant signal on the high energy line in $\sigma_{cross}$ configuration does not correspond to a trace of the larger $\sigma_{co}$ contribution that would not be perfectly rejected in the selection of polarization. 

At low excitation power, this weak PL appears at the center of a narrow hollow in the non-resonant PL background. At intermediate resonant power, a broader hollow appears around the resonance. The width and the depth of the hollow increase with the excitation power, it becomes asymmetric as it extends towards high energy (Fig.\ref{Fig8Sbis}). At the resonance, the PL intensity in $\sigma_{cross}$ does not increase significantly with the increase of the excitation power and at high power it is only slightly larger than the non-resonant background level (Fig.\ref{Fig8Sbis}). Such behavior was already observed in the time domain (see Fig.~\ref{Fig15S}) where the steady state PL intensity under a pulsed resonant excitation in $\sigma_{cross}$ was similar to the intensity obtained under CW non-resonant excitation. 

In $\sigma_{cross}$ configuration, one excites and detects the same Cr$^+$ spin state. In this case, the PL on line $1$ for an excitation on line $7$ results from a spin-flip of the electron with a conservation of the Cr$^+$ spin. Let's consider a $\sigma-$ resonant laser scanned around line $7$. It excites the spin state S$_z$=+5/2. A reduction observed in the non-resonant PL of line $1$ reflects a reduction of the occupation probability of the spin state S$_z$=+5/2. The broad hollow at high excitation power shows that a reduction of the occupation probability is obtained for an excitation far (meV range) from the resonance on line $7$.

The hollow in the non-resonant PL corresponds to an absorption in the acoustic phonon side-band of line $7$. At high resonant power, this weak absorption can induce an optical pumping which empties the states under excitation. This pumping results in a decrease of the non-resonant PL of line $1$. The depth of the absorption hollow increases with the resonant excitation power and it extends on the high energy side. The observed asymmetry corresponds to a better absorption on the high energy phonon side band of line $7$ ({\it i.e.} optical absorption with the emission of acoustic phonons). This is expected at low temperature where high energy phonons are not present for an optical absorption involving a simultaneous absorption of phonons ({\it i.e.} absorption on the low energy phonon side-band of line $7$). 

For such excitation in the phonon side-bands the reduction of the non-resonant PL is only partial and it does not reach zero. This shows that, in the presence of the additional non-resonant excitation, the optical pumping is not complete. Such partial destruction of the optical pumping by the non-resonant excitation was directly observed in the time domain in Fig.~\ref{Fig15S}.

At intermediate excitation power, in addition to the phonon side-band hollow, the resonant PL peak obtained for an excitation on the zero phonon line appears at the center of a small symmetric hollow. This narrow hollow is independent of the acoustic phonon band. It likely corresponds to an absorption in the pedestal of the zero-phonon line. This absorption increases when the laser gets closer to the resonance and enhances the optical pumping already observed for an excitation on the phonon side-band. At low excitation power, only the small symmetric hollow appears around the resonant absorption peak.

The surprising behavior is the increase of the PL for an excitation at the resonance on the zero-phonon line $7$. One could expect that the pumping would be further enhanced at the resonance and the amplitude of the non-resonant PL reduced. On the contrary, in this high resonant excitation regime, the resonant PL increases and returns to the level of the non-resonant background or slightly above. 

In this two excitation wavelengths configuration, the signal detected on line $1$ is the sum of the PL produced by the CW non-resonant excitation and the resonant PL. In $\sigma_{cross}$ the latter corresponds to a spin-flip of the electron with conservation of the spin of the Cr$^+$. A scan of the excitation energy around line $7$ corresponds to a modulation of the excitation power of this transition associated with the low energy hole-Cr$^+$ state. For an excitation in the phonon side band or on the pedestal of the zero-phonon line (i.e. weak excitation regime) we observe a decrease of the occupation probability of the resonantly excited hole-Cr$^+$ spin state. This reduction is induced by the optical pumping. At low excitation power, far from the saturation of the absorption, the pumping is expected to be linear with the absorption of the resonant laser. 

In the resonant power dependence of the PL excitation presented in Fig.~\ref{Fig8Sbis}, the total PL at the resonance in $\sigma_{cross}$ does not significantly depend on the increase of the resonant excitation power whereas the depth and the width of the hollow in the non-resonant PL increases. The depth of the hollow has also a sub-linear evolution with the increase of excitation power. This shows that at the resonance or close to the resonance, the saturation regime is reached ({\it i.e.} saturation of the absorption of the driven two levels system).

The observed increase of the PL of line $1$ at the resonance reflects a non-monotonic evolution of the PL with the increase of the excitation power on line $7$. It results from a modification of the spin dynamics under high intensity excitation. There are two possible sources for the modification of the dynamics. First, the spin-flip of the electron becomes not negligible at the resonance when the absorption is maximum and a high probability of occupation of the QD is reached. Secondly, the efficiency of the pumping of the low energy hole-Cr$^+$ state is reduced under high excitation intensity.

The dynamics of the electron is unlikely to be modified by the resonant excitation. On the contrary, a modification of the spin dynamics of the hole-Cr$^+$ induced by the generation of non-equilibrium acoustic phonon from the strongly driven QD is possible. Indeed a driven QD is an efficient local phonon source \cite{Harbush2010, Granger2012} although excited on resonance \cite{Poizat}. This happens during the formation of a polaron accompanied by the emission of acoustic wave packets \cite{Wigger2020}. This is especially the case in small II-VI QDs having a large potential deformation coupling of the exciton with acoustic phonon modes \cite{Besombes2001}. Phonons with an energy in the few 100 $\mu$eV range are emitted and are particularly adapted for direct single phonon spin-flip process among the hole-Cr$^+$ nano-magnet. A modeling of the impact of the polaron formation in a resonantly driven QD on the spin dynamics of an embedded magnetic atom is an interesting theoretical problem that goes beyond the scope of this experimental study.

\section{Conclusion}

To conclude, we have studied the resonant PL of the X$^+$-Cr$^+$ complex and the spin dynamics in Cr$^+$-doped CdTe/ZnTe QDs under combined resonant and non-resonant optical excitation. The resonant PL is  reduced by an optical pumping of the hole-Cr$^+$ spin. It can be partially restored by a high energy non-resonant excitation which erases the hole-Cr$^+$ spin memory.

An analysis of the energy distribution of the resonant PL shows that the spin dynamics of X$^+$-Cr$^+$ is dominated by spin tunneling processes. This is due to the weak energy splitting of this complex which limits its interaction with low energy acoustic phonons and to the absence of significant electron-Cr$^+$ exchange interaction. The spin dynamics in this system strongly differs from X$^{+}$-Mn$^{2+}$, the reference spin 5/2 system studied in the recent years. This is because of (i) the larger strain induced magnetic anisotropy for the Cr$^+$ spin and (ii) the very small exchange interaction of the Cr$^+$ spin with the electron spin.

Optical transitions that do not conserve the Cr$^+$ spin are observed in the resonant-PL excitation spectra. The energy shift of these transitions are controlled by the fine structure of the magnetic atom. This permits to determine the magnetic anisotropy of the Cr$^+$, in agreement with estimation obtained from the modelling of low magnetic field magneto-optic spectra.

Under optical excitation, a carrier-Cr$^+$ spin dynamics in a few tens of $ns$ range is observed in the auto-correlation of the non-resonant PL intensity, in the resonant optical pumping transients and in the non-resonant PL transient resulting from the hole-Cr$^+$ spin heating. In these three experiments, the observed dynamics is strongly sensitive to the optical excitation intensity. The effective spin temperature of the hole-Cr$^+$ system observed in the intensity distribution of the non-resonant PL is also significantly affected by the optical excitation conditions. This spin heating is induced by the generation of non-equilibrium acoustic phonons. A detailed analysis of the resonant PL suggests that the hole-Cr$^+$ spin dynamics can also be affected by phonons generated under resonant excitation. In future work, a modeling of the dynamics of the multi-level spin system in a Cr$^+$-doped QD and in particular its interaction with optically generated acoustic phonons will have to be performed.

\begin{acknowledgements}{}

The work was supported by the French ANR project MechaSpin (ANR-17-CE24-0024). V.T. acknowledges support from EU Marie Curie grant No 754303. The work in Tsukuba has been partially supported by JSPS Grants-in-Aid (KAKENH
I) for Challenging Research (Exploratory) (20K21116) and for Fostering Joint International Research (B) (20KK0113), JSPS Bilateral Joint Research Project (120219904). The collaboration between Univ. Grenoble Alpes and Univ. Tsukuba is performed in the framework of International Research Laboratories (IRL) "J-FAST" supported by CNRS. S.K. acknowledges the support from the Center for Spintronics Research Network (CSRN), Osaka University, Japan. 

\end{acknowledgements}


\begin{thebibliography}{}

\bibitem{Wolfowicz2021} G. Wolfowicz, F. J. Heremans, C. P. Anderson, S. Kanai, H. Seo, A. Gali, G. Galli,  D. D. Awschalom, Nature Reviews Materials 6, 906 (2021).

\bibitem{Besombes2004} L. Besombes, Y. Leger, L. Maingault, D. Ferrand, H. Mariette and J. Cibert, Phys. Rev. Lett. {\bf 93}, 207403 (2004).

\bibitem{Kudelski2007} A. Kudelski, A. Lemaitre, A. Miard, P. Voisin, T.C.M. Graham, R.J. Warburton and O. Krebs, Phys. Rev. Lett. {\bf 99}, 247209 (2007).

\bibitem{Kobak2014} J. Kobak, T. Smolenski, M. Goryca, M. Papaj, K. Gietka, A. Bogucki, M. Koperski, J.-G. Rousset, J. Suffczynski, E. Janik, M. Nawrocki, A. Golnik, P. Kossacki, W. Pacuski, Nature Com. {\bf 5}, 3191 (2014).

\bibitem{Besombes2012} L. Besombes, C.L. Cao, S. Jamet, H. Boukari, J. Fernandez-Rossier, Phys. Rev. B  {\bf 86}, 165306 (2012).

\bibitem{Krebs2013} O. Krebs, A. Lemaitre, Phys. Rev. Lett. {\bf 111}, 187401 (2013).

\bibitem{Varghese2014} B. Varghese, H. Boukari, L. Besombes, Phys. Rev. B {\bf 90}, 115307 (2014).

\bibitem{Bacher2016} R. Fainblat, C. J. Barrows, E.  Hopmann, S.  Siebeneicher, V.  A.  Vlaskin, D.  R. Gamelin, G.  Bacher, Nano Lett. {\bf 16}, 6371 (2016).

\bibitem{Bacher2020} S. Lorenz, C. S. Erickson, M. Riesner, D. R. Gamelin, R. Fainblat, G. Bacher, Nano Lett. 2020 {\bf 20}, 1896 (2020).

\bibitem{Bayer2020} E. V. Shornikova, D. R. Yakovlev, D. O. Tolmachev, V. Yu. Ivanov, I. V. Kalitukha, V. F. Sapega, D. Kudlacik, Y. G. Kusrayev, A. A. Golovatenko, S. Shendre, S. Delikanli, H. Volkan Demir, M. Bayer, ACS nano, {\bf 14}, 9032 (2020).

\bibitem{Besombes2014} L. Besombes, H. Boukari, Phys. Rev. B {\bf 89}, 085315 (2014).

\bibitem{Lafuente2017} A. Lafuente-Sampietro, H. Boukari, and L. Besombes, Phys.  Rev. B {\bf 95}, 245308 (2017).

\bibitem{Ciepla1975} M.Z. Cieplak, M. Godlewski, J.M. Baranowski, Phys. Stat. Sol. (b) {\bf 70}, 33, (1975).

\bibitem{Godlewski1980} M. Godlewski, J.M. Baranowski, Phys. Stat. Sol. (b) { \bf 97}, 281 (1980).

\bibitem{Tiwari2021} V. Tiwari, M. Arino, S. Gupta, M. Morita, T. Inoue, D. Caliste, P. Pochet, H. Boukari, S. Kuroda, L. Besombes, Phys. Rev. B {\bf 104}, L041301 (2021).

\bibitem{Lafuente2016} A. Lafuente-Sampietro, H. Utsumi, H. Boukari, S. Kuroda, L. Besombes, Phys. Rev. B {\bf 93}, 161301(R) (2016).

\bibitem{Lafuente2015} A. Lafuente-Sampietro, H. Boukari, L. Besombes, Phys. Rev. B {\bf 92}, 081305(R) (2015).


\bibitem{Ludwig1963} G.W. Ludwig and M.R. Lorenz, Phys. Rev. {\bf 131}, 601 (1963).

\bibitem{Vallin1974} J.T. Vallin, G.D. Watkins, Phys. Rev. B {\bf 9}, 2051 (1974).

\bibitem{Hawrylak2013} A. H. Trojnar, M. Korkusinski, U. C. Mendes, M. Goryca, M. Koperski, T. Smolenski, P. Kossacki, P. Wojnar, P. Hawrylak, Phys. Rev. B {\bf 87}, 205311 (2013).

\bibitem{Leger2007} Y. Leger, L. Besombes, L. Maingault, H. Mariette, Phys. Rev. B {\bf 76}, 045331 (2007).

\bibitem{Legall2012}  C. Le Gall, A. Brunetti, H. Boukari, and L. Besombes,  Phys. Rev. B, {\bf 85}, 195312 (2012).

\bibitem{Lafuente2018} A. Lafuente-Sampietro, H. Utsumi, M. Sunaga, K. Makita, H. Boukari, S. Kuroda, L. Besombes, Phys. Rev. B {\bf 97}, 155301 (2018).

\bibitem{Tiwari2020} V. Tiwari, K. Makita, M. Arino, S. Kuroda, H. Boukari, L. Besombes,  Phys. Rev. B {\bf 101}, 035305 (2020).

\bibitem{nanomagnet} D. Gatteschi, R. Sessoli and J. Villain, Molecular nanomagnet, Oxford University Press Inc., New York (2006).

\bibitem{Wernsdorfer2008} L. Bogani and W. Wernsdorfer, Nature Materials {\bf 7}, 179 (2008) and references therein.

\bibitem{Cao2011} C. L. Cao, L. Besombes, J. Fernandez-Rossier, Phys. Rev. B {\bf 84}, 205305 (2011).

\bibitem{Chud2005} E. M. Chudnovsky, D. A. Garanin, and R. Schilling, Phys. Rev. B {\bf 72}, 094426 (2005).

\bibitem{Tsit2005} E. Tsitsishvili, R. v. Baltz, and H. Kalt, Phys. Rev. B {\bf 72}, 155333 (2005).

\bibitem{Govorov2005} A. O. Govorov, A. V. Kalameitsev, Phys. Rev. B {\bf 71}, 035338 (2005).

\bibitem{Lafuente2016APL} A. Lafuente-Sampietro, H. Utsumi, H. Boukari, S. Kuroda, L. Besombes, Applied Physics Letters {\bf 109}, 053103 (2016).

\bibitem{Yakovlev2006} M. K. Kneip, D. R. Yakovlev, M. Bayer, A. A. Maksimov,
I. I. Tartakovskii, D. Keller, W. Ossau, L. W. Molenkamp, and
A. Waag, Phys. Rev. B 73, 035306 (2006).

\bibitem{Furdyna1988} J. K. Furdyna,  Journal of Applied Physics {\bf 64}, R29 (1988).

\bibitem{Abragam} A. Abragam  and B. Bleaney, Electron paramagnetic resonance of transition ions, Oxford University Press, Oxford (2012).

\bibitem{Besombes2001} L. Besombes, K. Kheng, L. Marsal, H. Mariette, Phys. Rev. B, {\bf 63}, 155307 (2001).

\bibitem{Harbush2010} D. Harbush, D. taubert, H.P. Tranitz, W. Wegscheider, S. Ludwig, Phys. Rev. Lett. {\bf 104}, 196801 (2010).

\bibitem{Granger2012} G. Granger, D. Taubert, C.E. Young, L. Gaudreau, A. Kam, S.A. Studenikin, P. Zawadzki, D. Harbush, D. Schuh, W. Wegscheider, Z.R. Wasilewski, A.A. Clerk, S. Ludwig, A.S. Sachrajda, Nature Physics {\bf 8}, 522 (2012).

\bibitem{Poizat} J. Kettler, N. Vaish, L. Mercier de Lepinay, B. Besga, P.-L. de assis, O. Bourgeois, A. Auffeve, m. Richard, J. Claudon, J.-M. Gerard, B. pigeau, O. Arcizet, P. Verlot, J.-P. Poizat, Nature Nanotechnology {\bf 16}, 283 (2021). 

\bibitem{Wigger2020} D. Wigger, V. Karakhanyan, C. Schneider, M. Kamp, S. Höfling, P. Machnikowski, T. Kuhn, and J. Kasprzak, Optics Letter {\bf 45}, 919 (2020).


\end{thebibliography}
\end{document}